%% file: main.tex
\documentclass[acmsmall,screen,nonacm]{acmart}

\input{preamble}
\input{macros}

\begin{document}

\title{\system: A Live Benchmark for Node.js Taint-Style Vulnerabilities}

\author{José Afonso}
\orcid{https://orcid.org/0009-0000-7288-398X}
\email{jose.pinto.afonso@tecnico.ulisboa.pt}
\affiliation{%
  \institution{INESC-ID / IST, ULisboa}
  \city{Lisbon}
  \country{Portugal}
}
\author{João Madeira Pereira}
\orcid{https://orcid.org/0009-0002-5404-719X}
\email{jhenriqu@andrew.cmu.edu}
\affiliation{%
  \institution{Carnegie Mellon University}
  \city{Pittsburgh}
  \country{USA}
}
\affiliation{%
  \institution{INESC-ID / IST, ULisboa}
  \city{Lisbon}
  \country{Portugal}
}
\author{Ronghao Ni}
\orcid{https://orcid.org/0009-0008-8518-5660}
\email{ronghaon@andrew.cmu.edu}
\affiliation{%
  \institution{Carnegie Mellon University}
  \city{Pittsburgh}
  \country{USA}
}
\author{Limin Jia}
\orcid{https://orcid.org/0000-0002-8160-349X}
\email{liminjia@andrew.cmu.edu}
\affiliation{%
  \institution{Carnegie Mellon University}
  \city{Pittsburgh}
  \country{USA}
}
\author{José Fragoso Santos}
\orcid{https://orcid.org/0000-0001-5077-300X}
\email{jose.fragoso.santos@tecnico.ulisboa.pt}
\affiliation{%
  \institution{INESC-ID / IST, ULisboa}
  \city{Lisbon}
  \country{Portugal}
}
\author{Pedro Adão}
\orcid{https://orcid.org/0000-0002-4049-1954}
\email{pedro.adao@tecnico.ulisboa.pt}
\affiliation{%
  \institution{Instituto de Telecomunicações / IST, ULisboa}
  \city{Lisbon}
  \country{Portugal}
}

\begin{abstract}
JavaScript vulnerabilities in npm packages pose a significant risk to the Node.js ecosystem, 
and a growing body of work aims to detect, exploit, and mitigate them. Studying and
evaluating this work requires up-to-date collections of real vulnerabilities, yet the most
commonly used datasets stop recording vulnerabilities in 2022.

We present \system, a live benchmark of real-world Node.js taint-style
vulnerabilities covering code and command injection, prototype
pollution, and path traversal. Built from \nEntries advisories collected through mid-2026, 
primarily from the GitHub Advisory Database, \system\ pairs each vulnerability with 
precise source and sink locations, a validated executable exploit and its
execution trace, and, when available, a patched version of the package.
\system supports long-term reproducibility and growth:
dependencies are mirrored in a self-hosted registry to guard against package
removal from the ecosystem, and dedicated infrastructure generates and validates exploits for newly reported vulnerabilities. The current release contains \nValidExploits validated exploits for
\nValidPackages packages. Using \system, we characterize how the frequency, 
structure, and complexity of Node.js taint-style vulnerabilities have evolved over time.
We find that these vulnerabilities are increasingly reported in large
packages: the median vulnerable package grew from roughly \medPkgSlocEarly
LoC in 2021--2022 to \medPkgSlocLate LoC in 2026. 
\end{abstract}




\maketitle

\input{Sections/intro-v1}

\input{Sections/motivation}

\input{Sections/methodology-v1}

\input{Sections/Dataset}

\input{Sections/Measurements}

\input{Sections/Discussion}

\input{Sections/Related_Work}

\input{Sections/Conclusion}

\bibliographystyle{ACM-Reference-Format}
\bibliography{references, references_llm_dataset}

\end{document}

%% file: preamble.tex
\usepackage{amsmath}

\usepackage{amssymb,amsfonts}
\usepackage{algorithmic}
\usepackage{graphicx}
\usepackage{array}
\usepackage{booktabs}
\usepackage{textcomp}
\usepackage{xcolor}

\usepackage{tikz}
\usetikzlibrary{arrows.meta,positioning,shapes.geometric,calc,fit,backgrounds,decorations.pathmorphing}

\usepackage{xspace}
\usepackage[inline]{enumitem}

\usepackage{listings}
\definecolor{lstkw}{HTML}{7B3FA0}    %
\definecolor{lstbi}{HTML}{1F7A8C}    %
\definecolor{lstcall}{HTML}{8A6400}  %
\definecolor{lststr}{HTML}{B0305A}   %
\definecolor{lstcmt}{HTML}{7A7A7A}   %
\definecolor{phsetup}{HTML}{2A78D6}
\definecolor{phexploit}{HTML}{D9531E}
\definecolor{phvalidate}{HTML}{0A8A4A}
\definecolor{phcleanup}{HTML}{8A8A85}
\lstdefinelanguage{JavaScript}{
  morekeywords={const,let,var,module,exports,async,await,function,
    return,delete,new,if,else,typeof,of,in},
  keywordstyle=\color{lstkw},
  morekeywords=[2]{setup}, keywordstyle=[2]\color{phsetup}\bfseries,
  morekeywords=[3]{exploit}, keywordstyle=[3]\color{phexploit}\bfseries,
  morekeywords=[4]{validate}, keywordstyle=[4]\color{phvalidate}\bfseries,
  morekeywords=[5]{cleanup}, keywordstyle=[5]\color{phcleanup}\bfseries,
  morekeywords=[6]{Object,console,assert,undefined,null,true,false,prototype,
    JSON,process,Promise,path},
  keywordstyle=[6]\color{lstbi},
  morekeywords=[7]{require,log,strictEqual,deepStrictEqual,ok,parse,stringify,
    resolve,dirname,run,execSync},
  keywordstyle=[7]\color{lstcall},
  sensitive=true,
  comment=[l]{//}, morecomment=[s]{/*}{*/},
  commentstyle=\color{lstcmt}\itshape,
  string=[b]", morestring=[b]',
  stringstyle=\color{lststr},
}
\lstdefinestyle{poc}{
  backgroundcolor=\color{black!3},
  frame=single, framerule=0.4pt, rulecolor=\color{black!22},
  numbers=left, numberstyle=\tiny\color{black!40}, numbersep=5pt,
  xleftmargin=1.5em, framexleftmargin=1.3em,
  framextopmargin=1pt, framexbottommargin=1pt,
  aboveskip=0pt, belowskip=0pt, breaklines=false,
  escapeinside={(*@}{@*)},
  emph={setPath}, emphstyle={\bfseries\color{phexploit}},
}
\newcommand{\pbar}[1]{\smash{\textcolor{#1}{\rule[-2.6pt]{2pt}{9.2pt}}}\hspace{4pt}}
\lstdefinestyle{pocplain}{style=poc, numbers=none, xleftmargin=0.3em,
  framexleftmargin=0.2em}

\def\BibTeX{{\rm B\kern-.05em{\sc i\kern-.025em b}\kern-.08em
    T\kern-.1667em\lower.7ex\hbox{E}\kern-.125emX}}

\newcommand{\system}{\textsc{x.js}\xspace}
\newcommand{\vulcan}{\textsc{VulcaN}\xspace}
\newcommand{\secbench}{\textsc{SecBench.js}\xspace}

\newcommand{\cstep}[1]{\tikz[baseline=(c.base)]\node[circle, draw=black, fill=white,
  text=black, font=\scriptsize\bfseries, inner sep=0pt, minimum size=3.4mm]
  (c) {#1};}

\DeclareRobustCommand{\capyes}{\textcolor[HTML]{C0392B}{\Large$\bullet$}}
\DeclareRobustCommand{\capno}{\textcolor{black!35}{\Large$\circ$}}

\newcommand{\runin}[1]{\emph{#1.}\hspace{0.5em}}

\newcommand{\takeaway}[1]{\par\smallskip\noindent
  \colorbox{gray!12}{\parbox{\dimexpr\linewidth-2\fboxsep\relax}{\textbf{Takeaway.} #1}}\par\smallskip}

%% file: macros.tex
\newif\ifdraftnumbers
\draftnumbersfalse
\DeclareRobustCommand{\dyn}[1]{\ifdraftnumbers\textcolor{red}{#1}\else#1\fi}

\newcommand{\nGhsaAdvisories}{\dyn{1{,}466}\xspace} %

\newcommand{\nSnykLinked}{\dyn{1{,}357}\xspace}     %
\newcommand{\nOsvSnykRefs}{\dyn{339}\xspace}        %
\newcommand{\pctPocOnlySnyk}{\dyn{10.4\%}\xspace}   %
\newcommand{\pctPocOnlyGhsa}{\dyn{8.5\%}\xspace}    %

\newcommand{\nAddedTotal}{\dyn{272}\xspace}
\newcommand{\nAddedVulcan}{\dyn{69}\xspace}
\newcommand{\nAddedSecbench}{\dyn{203}\xspace}
\newcommand{\nAddedWrongCwe}{\dyn{111}\xspace}      %
\newcommand{\nAddedWithoutCwe}{\dyn{8}\xspace}      %
\newcommand{\nAddedSnykOnly}{\dyn{153}\xspace}      %

\newcommand{\nEntries}{\dyn{1{,}738}\xspace}
\newcommand{\nEntriesCve}{\dyn{1{,}428}\xspace}
\newcommand{\nEntriesGhsa}{\dyn{1{,}585}\xspace}
\newcommand{\nEntriesSnyk}{\dyn{1{,}604}\xspace}

\newcommand{\nValidExploits}{\dyn{1{,}159}\xspace}
\newcommand{\nValidPackages}{\dyn{965}\xspace}
\newcommand{\pctFixFromAdvisory}{\dyn{65{.}4\%}\xspace}  %

\newcommand{\medPkgSlocEarly}{\dyn{250}\xspace}   %
\newcommand{\medPkgSlocLate}{\dyn{6{,}100}\xspace}   %

\newcommand{\nPPAdv}{\dyn{534}\xspace}              %
\newcommand{\nPPPkgs}{\dyn{418}\xspace}             %
\newcommand{\nPPFlow}{\dyn{460}\xspace}             %
\newcommand{\nPPCapab}{\dyn{448}\xspace}            %
\newcommand{\nPPPatches}{\dyn{375}\xspace}          %
\newcommand{\nPPPatchValid}{\dyn{329}\xspace}       %
\newcommand{\nPPPatchInvalid}{\dyn{46}\xspace}      %
\newcommand{\pctPPFlow}{\dyn{86\%}\xspace}          %
\newcommand{\pctPPCapab}{\dyn{84\%}\xspace}         %
\newcommand{\pctPPPatchValid}{\dyn{88\%}\xspace}    %
\newcommand{\pctPPPatchInvalid}{\dyn{12\%}\xspace}   %
\newcommand{\nPTAdv}{\dyn{525}\xspace}              %
\newcommand{\nPTPkgs}{\dyn{407}\xspace}             %
\newcommand{\nPTFlow}{\dyn{405}\xspace}             %
\newcommand{\nPTCapab}{\dyn{359}\xspace}            %
\newcommand{\nPTPatches}{\dyn{256}\xspace}          %
\newcommand{\nPTPatchValid}{\dyn{224}\xspace}       %
\newcommand{\nPTPatchInvalid}{\dyn{32}\xspace}      %
\newcommand{\pctPTFlow}{\dyn{77\%}\xspace}          %
\newcommand{\pctPTCapab}{\dyn{68\%}\xspace}         %
\newcommand{\pctPTPatchValid}{\dyn{88\%}\xspace}    %
\newcommand{\pctPTPatchInvalid}{\dyn{12\%}\xspace}   %
\newcommand{\nCmdIAdv}{\dyn{354}\xspace}            %
\newcommand{\nCmdIPkgs}{\dyn{274}\xspace}           %
\newcommand{\nCmdIFlow}{\dyn{274}\xspace}           %
\newcommand{\nCmdICapab}{\dyn{258}\xspace}          %
\newcommand{\nCmdIPatches}{\dyn{180}\xspace}        %
\newcommand{\nCmdIPatchValid}{\dyn{151}\xspace}     %
\newcommand{\nCmdIPatchInvalid}{\dyn{29}\xspace}    %
\newcommand{\pctCmdIFlow}{\dyn{77\%}\xspace}        %
\newcommand{\pctCmdICapab}{\dyn{73\%}\xspace}       %
\newcommand{\pctCmdIPatchValid}{\dyn{84\%}\xspace}  %
\newcommand{\pctCmdIPatchInvalid}{\dyn{16\%}\xspace}%
\newcommand{\nCodeIAdv}{\dyn{203}\xspace}           %
\newcommand{\nCodeIPkgs}{\dyn{144}\xspace}          %
\newcommand{\nCodeIFlow}{\dyn{101}\xspace}           %
\newcommand{\nCodeICapab}{\dyn{94}\xspace}          %
\newcommand{\nCodeIPatches}{\dyn{136}\xspace}       %
\newcommand{\nCodeIPatchValid}{\dyn{108}\xspace}    %
\newcommand{\nCodeIPatchInvalid}{\dyn{28}\xspace}   %
\newcommand{\pctCodeIFlow}{\dyn{50\%}\xspace}       %
\newcommand{\pctCodeICapab}{\dyn{46\%}\xspace}      %
\newcommand{\pctCodeIPatchValid}{\dyn{79\%}\xspace} %
\newcommand{\pctCodeIPatchInvalid}{\dyn{21\%}\xspace}%
\newcommand{\nAllAdv}{\dyn{1{,}616}\xspace}         %
\newcommand{\nAllPkgs}{\dyn{1{,}173}\xspace}        %
\newcommand{\nAllFlow}{\dyn{1{,}240}\xspace}        %
\newcommand{\nAllCapab}{\dyn{1{,}159}\xspace}       %
\newcommand{\nNoFlow}{\dyn{376}\xspace}             %
\newcommand{\nNoCapability}{\dyn{81}\xspace}        %
\newcommand{\nAllPatches}{\dyn{947}\xspace}     %
\newcommand{\nAllPatchValid}{\dyn{812}\xspace}      %
\newcommand{\nAllPatchInvalid}{\dyn{135}\xspace}     %
\newcommand{\pctAllFlow}{\dyn{77\%}\xspace}         %
\newcommand{\pctAllCapab}{\dyn{72\%}\xspace}        %
\newcommand{\pctAllPatchValid}{\dyn{86\%}\xspace}   %
\newcommand{\pctAllPatchInvalid}{\dyn{14\%}\xspace}  %
\newcommand{\nAllPocs}{\dyn{1{,}694}\xspace}        %

\newcommand{\nAdvNew}{\dyn{1{,}041}\xspace}          %
\newcommand{\nAdvFromSecBench}{\dyn{464}\xspace}     %
\newcommand{\nAdvFromVulcan}{\dyn{243}\xspace}       %
\newcommand{\nAdvBoth}{\dyn{132}\xspace}          %
\newcommand{\nExclUnpublished}{\dyn{37}\xspace}     %
\newcommand{\nExclGenerated}{\dyn{74}\xspace}        %
\newcommand{\nExclInstall}{\dyn{5}\xspace}         %
\newcommand{\nExclGui}{\dyn{6}\xspace}             %
\newcommand{\nExcl}{\dyn{122}\xspace}              %
\newcommand{\nPPCapOnly}{\dyn{497}\xspace}        %
\newcommand{\nPTCapOnly}{\dyn{441}\xspace}        %
\newcommand{\nCmdICapOnly}{\dyn{300}\xspace}      %
\newcommand{\nCodeICapOnly}{\dyn{159}\xspace}     %
\newcommand{\nAllCapOnly}{\dyn{1{,}397}\xspace}    %
\newcommand{\nCapOnlyUndated}{\dyn{7}\xspace}    %
\newcommand{\pctPPCapOnly}{\dyn{93\%}\xspace}     %
\newcommand{\pctPTCapOnly}{\dyn{84\%}\xspace}     %
\newcommand{\pctCmdICapOnly}{\dyn{85\%}\xspace}   %
\newcommand{\pctCodeICapOnly}{\dyn{78\%}\xspace}  %
\newcommand{\pctAllCapOnly}{\dyn{86\%}\xspace}    %
\newcommand{\nTrendAdvTwentySix}{\dyn{278}\xspace}   %
\newcommand{\pctTopFiveTwentyOne}{\dyn{7\%}\xspace}   %
\newcommand{\pctTopFiveTwentySix}{\dyn{32\%}\xspace}   %
\newcommand{\nOpenclawTwentySix}{\dyn{56}\xspace}   %
\newcommand{\medPkgSlocEighteen}{\dyn{75}\xspace}   %
\newcommand{\medPkgSlocTwentySix}{\dyn{20{,}600}\xspace}   %
\newcommand{\medSetupEarly}{\dyn{7--8}\xspace}   %
\newcommand{\medSetupLate}{\dyn{27}\xspace}   %
\newcommand{\medExploitEarly}{\dyn{3--4}\xspace}   %
\newcommand{\medExploitLate}{\dyn{7}\xspace}   %
\newcommand{\medExecEarly}{\dyn{18--21}\xspace}   %
\newcommand{\medExecLate}{\dyn{63}\xspace}   %
\newcommand{\pctNoFixEighteen}{\dyn{70\%}\xspace}   %
\newcommand{\pctNoFixTwentyOne}{\dyn{44\%}\xspace}   %
\newcommand{\pctNoFixTwentySix}{\dyn{6\%}\xspace}   %
\newcommand{\pctShortBeforeTwentyOne}{\dyn{41\%}\xspace}   %
\newcommand{\pctShortBeforeTwentySix}{\dyn{90\%}\xspace}   %
\newcommand{\pctAfterMax}{\dyn{17\%}\xspace}   %
\newcommand{\medChurnEarly}{\dyn{17--48}\xspace}   %
\newcommand{\medChurnTwentyFive}{\dyn{70}\xspace}   %
\newcommand{\medChurnTwentySix}{\dyn{235}\xspace}   %
\newcommand{\medChurnTwentySixPkg}{\dyn{112}\xspace}   %
\newcommand{\nAiTwentyFive}{\dyn{3}\xspace}   %
\newcommand{\nAiTwentySix}{\dyn{12}\xspace}   %
\newcommand{\nPatchedTwentySix}{\dyn{245}\xspace}   %
\newcommand{\medChurnAi}{\dyn{160}\xspace}   %
\newcommand{\medChurnNoAi}{\dyn{162}\xspace}   %
\newcommand{\nRecurTwentySix}{\dyn{125}\xspace}   %
\newcommand{\pctRecurTwentySix}{\dyn{45\%}\xspace}   %
\newcommand{\pctRecurTwentySixNoOpenclaw}{\dyn{34\%}\xspace}   %
\newcommand{\medPkgSlocTwentySixNoOpenclaw}{\dyn{10{,}900}\xspace}   %
\newcommand{\medSetupTwentySixPkg}{\dyn{21}\xspace}   %
\newcommand{\pctRecurEarlyRange}{\dyn{1--10\%}\xspace}   %
\newcommand{\pctRecurTwentyFive}{\dyn{14\%}\xspace}   %
\newcommand{\nRecurOpenclaw}{\dyn{50}\xspace}   %
\newcommand{\pctPkgRecurEarly}{\dyn{10\%}\xspace}   %
\newcommand{\pctPkgRecurTwentySix}{\dyn{21\%}\xspace}   %
\newcommand{\medLagEighteen}{\dyn{162}\xspace}   %
\newcommand{\medLagTwentyOne}{\dyn{64}\xspace}   %
\newcommand{\medLagRecent}{\dyn{1--5}\xspace}   %
\newcommand{\nTrendAdvPreEighteen}{\dyn{26}\xspace}   %
\newcommand{\medPkgSlocEarlyAdv}{\dyn{260--290}\xspace}   %
\newcommand{\medPkgSlocTwentyFourAdv}{\dyn{2{,}400}\xspace}   %
\newcommand{\medExecLatePkg}{\dyn{47}\xspace}   %
\newcommand{\nGapRecur}{\dyn{139}\xspace}   %
\newcommand{\nGapSameDay}{\dyn{9}\xspace}   %
\newcommand{\nGapAI}{\dyn{61}\xspace}   %
\newcommand{\medGapAI}{\dyn{11}\xspace}   %
\newcommand{\pctGapAIMonth}{\dyn{80\%}\xspace}   %
\newcommand{\nGapOther}{\dyn{42}\xspace}   %
\newcommand{\nGapOtherPkgs}{\dyn{16}\xspace}   %
\newcommand{\medGapOtherYears}{\dyn{2.8}\xspace}   %
\newcommand{\pctGapOtherYear}{\dyn{60\%}\xspace}   %
\newcommand{\nGapOtherLong}{\dyn{25}\xspace}   %
\newcommand{\nGapOtherLongSpring}{\dyn{20}\xspace}   %
\newcommand{\nAdvMultiEntry}{\dyn{78}\xspace}   %
\newcommand{\nSampleValid}{\dyn{100}\xspace}     %
\newcommand{\nSampleNoFlow}{\dyn{30}\xspace}     %
\newcommand{\nSampleInvalid}{\dyn{10}\xspace}    %
\newcommand{\nNoFlowTimeout}{\dyn{16}\xspace}
\newcommand{\nNoFlowNative}{\dyn{7}\xspace}
\newcommand{\nNoFlowCrash}{\dyn{3}\xspace}
\newcommand{\nNoFlowCwe}{\dyn{3}\xspace}
\newcommand{\nNoFlowSink}{\dyn{1}\xspace}
\newcommand{\nInvalidTrue}{\dyn{9}\xspace}
\newcommand{\pctPkgGapTwentyOne}{\dyn{6\%}\xspace}   %
\newcommand{\pctPkgGapTwentyTwo}{\dyn{14\%}\xspace}   %
\newcommand{\pctPkgGapTwentyFour}{\dyn{7\%}\xspace}   %
\newcommand{\pctNoOpenclawDropTwentySix}{\dyn{47\%}\xspace}   %
\newcommand{\pctPerPkgDropTwentySix}{\dyn{70\%}\xspace}   %
\newcommand{\nPatchNoBlock}{\dyn{94}\xspace}   %

%% file: Sections/intro-v1.tex
\section{Introduction}
\label{sec:intro}

JavaScript is among the most widely used programming languages, and much of its reach today comes from Node.js~\cite{nodejs}, a V8-based server-side runtime. Node.js underpins production software ranging from developer tooling~\cite{vscode} to widely deployed desktop applications~\cite{discord,slack}; its package ecosystem, npm, is the world's largest software repository, with millions of reusable packages.
 Vulnerabilities in npm packages
threaten both their direct users and the wider ecosystem: a single flaw
can affect thousands of downstream packages through their dependencies,
potentially enabling sophisticated supply-chain
attacks~\cite{liu2022demystifying}. This risk has motivated substantial
recent work on detecting, confirming, and repairing server-side
JavaScript vulnerabilities~\cite{cassel2025nodemedicfine,cassel2023nodemedic,cornelissen2024ghunter,ferreira2024graphjs,kang2023FAST,li2022odgen,staicu2018synode}.
LLM-based tools have also been applied to vulnerability detection and confirmation~\cite{llmvdjs,vulrag,nguyen2025safe,ullah2024llms,10.1145/3808132,li2025iris,simsek2026pocgen}.

Most tools focus on vulnerabilities with severe consequences~\cite{li2022odgen,kang2023FAST,cassel2023nodemedic,ferreira2024graphjs,marques2025automated}: 
code injection, in which attacker-controlled input is evaluated as program code; OS command injection, in which it reaches an operating system command; prototype pollution, in which it modifies the attributes of a built-in object prototype; and path traversal, in which it manipulates a file-system path to access unintended files.
Prior work~\cite{cassel2023nodemedic,ferreira2024graphjs} refers to these as \emph{taint-style} vulnerabilities as they occur when attacker-controlled (\emph{tainted}) inputs enter a program through its public APIs, known as \emph{sources}, and propagate to sensitive code locations, known as \emph{sinks}, such as \texttt{eval()}.

Evaluating vulnerability-detection tools rigorously requires carefully engineered ground-truth datasets. Such datasets must be accurate, containing only true vulnerabilities, and rich in metadata to support vulnerability validation and deduplication. 
For taint-style vulnerabilities, desirable metadata includes a proof-of-concept (PoC) exploit for confirming the vulnerability, the locations of the sources and sinks, and the PoC's execution trace, which helps uniquely identify a vulnerability instance. 
Further, recent vulnerability datasets~\cite{lee2025secbench,zhang2026bountybench,nie2026secodeplt,wang2026cybergym,primevul} support the evaluation of code LLMs and agents, and have introduced paired vulnerable and patched versions of functions or packages to assess LLM hallucinations and their tendency to over-report vulnerabilities. 
The two most commonly used datasets for Node.js taint-style vulnerabilities are \vulcan~\cite{brito2023study} and \secbench~\cite{bhuiyan2023secbench}. 
However, both datasets have been shown to contain inaccurate and incomplete vulnerability information~\cite{ferreira2024graphjs}, lack validated patches, and most importantly, are outdated, with the most recent CVE dating back to 2022. Recent trends, such as the widespread adoption of code LLMs, may have reshaped the software vulnerability landscape, raising concerns about the representativeness of evaluations based on older packages. Notably, vulnerabilities reported in recent years affect substantially larger packages, with the median vulnerable package size increasing from approximately \medPkgSlocEarly LoC in 2021--2022 to \medPkgSlocLate LoC in 2026 (\S\ref{sec:metrics}).

To fill this gap and enable rigorous and reproducible tool evaluations, we present \system, a live benchmark of Node.js taint-style vulnerabilities, supported by a pipeline for advisory collection and PoC validation. \system is designed to grow and to last.
\system's collection and validation pipeline turns newly reported advisories 
in the GitHub Advisory Database~\cite{ghsa} into entries in the dataset. 
Then an LLM agent writes a candidate PoC, which is then validated. 
Manual inspection is needed only when a candidate PoC fails validation.
To keep entries executable as npm packages are updated, unpublished, or removed, \system uses a self-hosted package mirror, backed by Verdaccio~\cite{verdaccio}, that stores the exact package versions each entry requires.
Finally,
each entry is paired, when one exists, with a patched version that 
has been checked against the~PoC.

The main challenge is to generate the required metadata automatically
and at scale while maintaining high precision. This requires a clear
definition of a \emph{valid} PoC and tools that reliably identify and
reject invalid ones. Collecting precise source and sink locations
and execution traces also poses a scalability challenge as naive
instrumentation does not scale to large packages. \system uses
on-demand instrumentation to collect this information at scale.
The current release covers advisories in the four targeted taint-style
vulnerability classes from the GitHub Advisory Database~\cite{ghsa}
up to mid-2026, with links to the corresponding Snyk~\cite{snyk}
advisories, and contains \nValidExploits validated exploits for \nValidPackages packages:
more than \dyn{twice} as many vulnerabilities as \secbench and more than
\dyn{three} times as many as \vulcan in the same four classes. 
Leveraging \system, we examine how Node.js vulnerabilities have evolved over time. We characterize their complexity through the size of vulnerable packages and the amount of package and exploit code executed, and investigate how often a package is affected again by the same vulnerability class after it has been patched. Finally, we find two packages whose latest releases remain exploitable although their advisories list them as patched.
In summary, this paper makes the following contributions:
\begin{itemize}[leftmargin=*]
  \item A definition of exploit validity for taint-style vulnerabilities and a set of dynamic analyses that check it automatically, recovering the source, sink, and execution trace of each vulnerability along the way.
  \item \system, a live benchmark that currently contains \nValidExploits Node.js taint-style vulnerabilities across code injection, command injection, prototype pollution, and path traversal, each demonstrated by an executable exploit and, when one exists, paired with a patch checked against it.
  \item A pipeline that makes \system live: it incorporates newly reported advisories, generates and validates their exploits, checks their patches, and keeps entries reproducible through a self-hosted mirror of all required package versions.
  \item An empirical study, enabled by \system, of how the complexity of Node.js taint-style vulnerabilities has evolved over time and how often the same vulnerability class recurs in a package after being~patched.
\end{itemize}

The remainder of this paper is organized as follows.
Section~\ref{sec:motivation} presents the background and motivation, including our attacker model, proof-of-concept exploits, and the conditions for exploit validity.
Section~\ref{sec:methodology} describes our pipeline for   collecting advisories, constructing and validating exploits, checking patches, and gathering metadata.
Section~\ref{sec:benchmark} presents \system, its contents, its comparison with existing benchmarks, and how its entries remain executable over time.
Section~\ref{sec:metrics} uses \system to study how Node.js vulnerabilities have evolved.
Section~\ref{sec:discussion} discusses limitations and threats to validity.
Section~\ref{sec:relatedWork} surveys related work, and Section~\ref{sec:conclusionsAndFutureWork} concludes.

%% file: Sections/motivation.tex
\section{Background and Motivation}
\label{sec:motivation}

We explain the attacker model (\S\ref{subsec:threat-model}), introduce proof-of-concept exploits and their structure (\S\ref{subsec:poc-structure}), and discuss the conditions for exploit validity and the challenges of automated validation (\S\ref{ssec:motivation-poc-validation}).

\subsection{Attacker model}
\label{subsec:threat-model}
\system\ targets four vulnerability classes: path traversal
(CWE-22), OS command injection (CWE-78), code injection (CWE-94), and
prototype pollution (CWE-1321). We adopt the threat model from prior work on Node.js vulnerability analysis~\cite{staicu2018synode,li2022odgen,cassel2023nodemedic}.
We assume an attacker who interacts with a vulnerable package only
through its \emph{public interface}, i.e., the set of functions and objects it explicitly exports.
We call the propagation of attacker-controlled data from sources to sinks a tainted \emph{flow}.
We further require that the attacker cannot exercise the following capabilities except through the package’s public interface:
\begin{itemize}[leftmargin=*]
  \item \emph{reach package internals directly}: they cannot call
  non-exported functions nor modify the package's internal state, such as
  module-level variables that persist across calls, other than through the
  public interface;
  \item \emph{act on the execution environment}: they cannot create nor modify
  files, spawn processes, nor change the system configuration; and
  \item \emph{execute arbitrary code or modify global state}: they cannot evaluate code or modify global objects or built-in prototypes such as \texttt{Object.prototype}.
\end{itemize}
The latter two restrictions exclude capabilities that would render the vulnerability irrelevant. For example, an attacker who can already write files, execute commands, or run arbitrary code has no need to exploit a package to obtain those capabilities.

\subsection{Proof-of-Concept Exploit (PoC)}
\label{subsec:poc-structure}
\begin{figure}[t]
  \centering
  \begin{minipage}[c]{0.445\textwidth}
    \input{Listings/free-form-poc-section2.tex}
  \end{minipage}\hfill
  \begin{minipage}[c]{0.04\textwidth}
    \centering
    \tikz\draw[-{Stealth[length=2.4mm]}, line width=1pt, black!40] (0,0) -- (0.45,0);
  \end{minipage}\hfill
  \begin{minipage}[c]{0.50\textwidth}
    \input{Listings/lifecycle-poc-section2.tex}
  \end{minipage}
  \caption{Code on the left is a free-form
  PoC. It is translated into \system's lifecycle format on the right. The
  colored bars map every line to its phase:
  \textcolor{phsetup}{\textbf{setup}},
  \textcolor{phexploit}{\textbf{exploit}},
  \textcolor{phvalidate}{\textbf{validate}}, and
  \textcolor{phcleanup}{\textbf{cleanup}}.}
  \label{fig:poc-format}
  \Description{Two JavaScript listings side by side. Left: the four-line
  running example, which loads vuln-pkg, prints an unset property of an
  empty object, calls setPath with the path __proto__.polluted, and prints the
  property again, now set; each line carries a colored bar for its phase.
  Right: the same exploit as a module with four async functions: setup
  deletes Object.prototype.polluted, loads the package and asserts the
  property is undefined; exploit calls setPath; validate asserts the property
  equals "yes"; cleanup deletes it again.}
  \vspace{-10pt}
\end{figure}
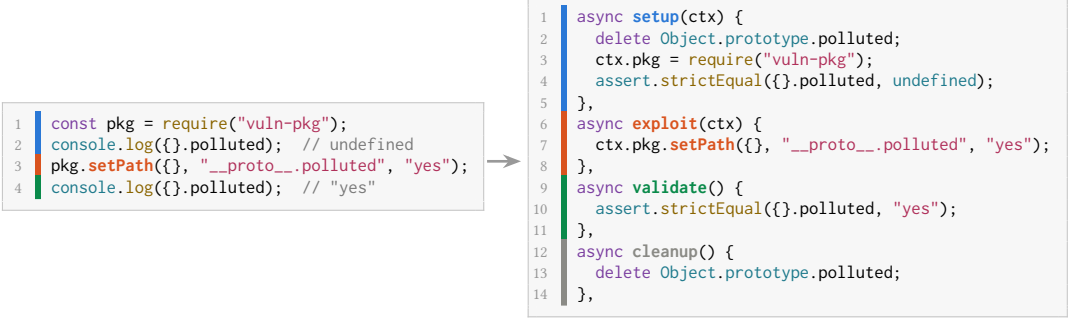

\paragraph{PoC formats.} A PoC is a program that, when executed together with a vulnerable package, demonstrates a feasible exploitation path. 
To enable automated validation, prior works design PoCs to produce observable side effects; for example, a command-injection PoC may create a file with a specific file name to signal successful exploitation~\cite{cassel2025nodemedicfine,marques2025automated,simsek2026pocgen}; %
whereas a prototype-pollution PoC (left of Figure~\ref{fig:poc-format}) may call the vulnerable function with a path that walks from the target object to \texttt{Object.prototype} (line~\ref{lst:ex-attack}), and then check that the property is present on a fresh object (line~\ref{lst:ex-check}).

\system\ distinguishes two forms of PoC, illustrated in Figure~\ref{fig:poc-format}. 
A free-form PoC directly triggers the vulnerability (Figure~\ref{fig:poc-format}, left).
PoCs that typically accompany advisories, as well as those in prior research datasets, are free-form.
However, free-form PoCs are ill-suited to automated validation: they interleave setup code and attacker actions, leave preconditions implicit, and often signal success through console output, making it hard to tell whether the effect was produced by the package. 
In contrast, a \emph{lifecycle} PoC (Figure~\ref{fig:poc-format}, right) splits the same exploit into four distinct phases: 
\textcolor{phsetup}{\texttt{setup}} establishes the preconditions;
\textcolor{phexploit}{\texttt{exploit}} runs the payload through the package;
\textcolor{phvalidate}{\texttt{validate}} checks the resulting effect; and 
\textcolor{phcleanup}{\texttt{cleanup}} restores
the environment. 
The lifecycle format explicitly distinguishes between harness and attacker's actions: \textcolor{phsetup}{\texttt{setup}}, \textcolor{phvalidate}{\texttt{validate}}, and \textcolor{phcleanup}{\texttt{cleanup}} functions
belong to the harness, while the \textcolor{phexploit}{\texttt{exploit}} function is the attack. 
\system\ stores PoCs in lifecycle format.

\subsection{PoC validation}
\label{ssec:motivation-poc-validation}
Observing the intended side effect does not establish vulnerability: a PoC may produce the effect itself or assert a condition that holds regardless of exploitation. This concern is particularly acute for LLM-generated exploits, which may game such checks through reward-hacking behaviors~\cite{simsek2026pocgen,zhong2026impossiblebench,baker2025monitoring}.
Based on the attacker model, we consider a PoC to be \emph{valid} only if it satisfies the following conditions:
(i)~attacker-controlled data flows from a source of the package to a sink of the reported class; (ii)~the attack produces an observable effect that is confirmed by a subsequent check; and (iii)~the exploit does not use any capability that would allow it to produce that effect without the package.

\paragraph{Challenges of dynamic analysis.} 
To check conditions~(i) and~(ii), \system uses dynamic taint analyses, which trace attacker (tainted) inputs to the targeted sink calls. 
Applying these analyses to real packages raises the challenge of deciding how much of the package and its dependencies to instrument.
Real-world packages often have large dependency trees, making full instrumentation prohibitively expensive, while limited instrumentation can lead to taint loss in uninstrumented code. 
To address these challenges, \system employs an on-demand instrumentation scheme that improves the performance of dynamic taint analysis while preserving soundness (\S\ref{sec:methodology}). %

\paragraph{Challenges of capability analysis.}
Enforcing condition (iii) raises the challenge of \emph{attribution}:
determining whether the observed side effect results from exploiting
the vulnerability or from the PoC's harness.
A harness may legitimately perform privileged operations, such as
creating a temporary directory for test inputs and outputs, but it
must not produce the effect itself.
For example, a command-injection PoC whose harness calls
\texttt{child\_process.exec} directly executes a command without
going through the package, and is therefore invalid.
To address this challenge, we develop a capability-checking system
that determines whether the capabilities gained and exercised by a
PoC are attributable to exploitation of the vulnerability or to
other code, such as the harness.
Our lifecycle PoC format supports this attribution, as detailed
in \S\ref{sec:methodology}.

%% file: Listings/free-form-poc-section2.tex
\begin{lstlisting}[style=poc]
(*@\pbar{phsetup}@*)const pkg = require("vuln-pkg");
(*@\pbar{phsetup}@*)console.log({}.polluted);  // undefined(*@\label{lst:ex-before}@*)
(*@\pbar{phexploit}@*)pkg.setPath({}, "__proto__.polluted", "yes");(*@\label{lst:ex-attack}@*)
(*@\pbar{phvalidate}@*)console.log({}.polluted);  // "yes"(*@\label{lst:ex-check}@*)
\end{lstlisting}

%% file: Listings/lifecycle-poc-section2.tex
\begin{lstlisting}[style=poc]
(*@\pbar{phsetup}@*)async setup(ctx) {
(*@\pbar{phsetup}@*)  delete Object.prototype.polluted;(*@\label{lst:poc-reset}@*)
(*@\pbar{phsetup}@*)  ctx.pkg = require("vuln-pkg");
(*@\pbar{phsetup}@*)  assert.strictEqual({}.polluted, undefined);
(*@\pbar{phsetup}@*)},
(*@\pbar{phexploit}@*)async exploit(ctx) {
(*@\pbar{phexploit}@*)  ctx.pkg.setPath({}, "__proto__.polluted", "yes");(*@\label{lst:poc-attack}@*)
(*@\pbar{phexploit}@*)},
(*@\pbar{phvalidate}@*)async validate() {
(*@\pbar{phvalidate}@*)  assert.strictEqual({}.polluted, "yes");
(*@\pbar{phvalidate}@*)},
(*@\pbar{phcleanup}@*)async cleanup() {
(*@\pbar{phcleanup}@*)  delete Object.prototype.polluted;
(*@\pbar{phcleanup}@*)},
\end{lstlisting}

%% file: Sections/methodology-v1.tex
\section{Methodology}
\label{sec:methodology}

We build \system\ by collecting vulnerability advisories and turning
candidate exploits into reproducible evidence of vulnerable behavior.
This section describes how we collect advisories
(\S\ref{subsec:collection}), presents the pipeline that constructs and
validates exploits and checks their patches (\S\ref{subsec:pipeline}), and
details our approach to PoC validation (\S\ref{subsec:validation}).

\begin{figure}[t]
  \centering
  \includegraphics[scale=0.9]{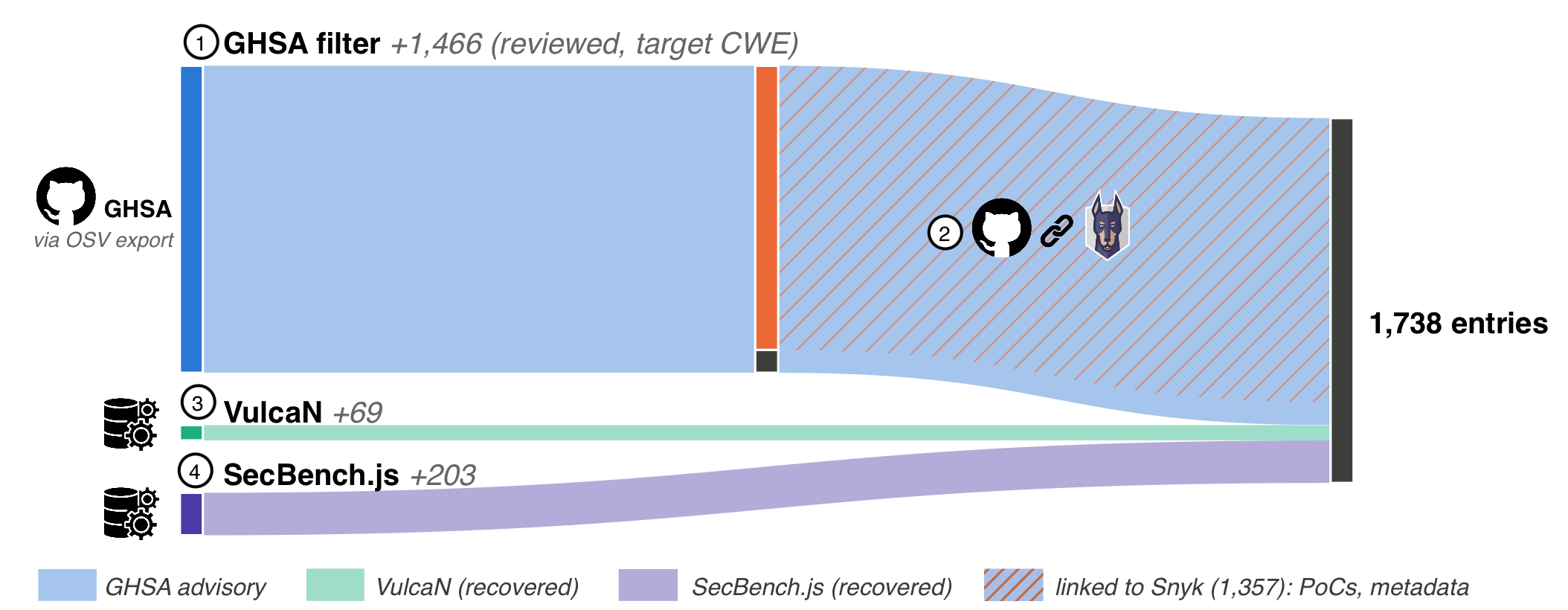}
  \caption{Advisory collection. Band widths are proportional to the number
  of advisories.}
  \Description{A Sankey diagram of advisory collection. GHSA, accessed through
  the OSV bulk export and filtered to reviewed, non-withdrawn advisories in the
  four target CWE classes, contributes \nGhsaAdvisories advisories; \nSnykLinked
  of them are linked to a Snyk advisory, which enriches them with
  proof-of-concept exploits and metadata without adding entries. VulcaN adds
  \nAddedVulcan and SecBench.js adds \nAddedSecbench further vulnerabilities,
  giving \nEntries entries in total.}
  \label{fig:collection-overview}
\end{figure}

\subsection{Advisory Collection}
\label{subsec:collection}

Figure~\ref{fig:collection-overview}
summarizes how we collect advisories for our target vulnerability classes from four sources.
We retrieve all npm advisories from the GitHub Advisory Database
(GHSA)~\cite{ghsa} using the Open Source Vulnerabilities
(OSV) service's bulk export~\cite{osv}~\cstep{1}.
We discard withdrawn and unreviewed advisories, which have
not been assessed by GitHub for quality and often
lack the affected-version ranges needed to install a vulnerable
version~\cite{ghsa-docs}.
We enrich the remaining GHSA advisories with Snyk
metadata~\cite{snyk}~\cstep{2}, particularly PoCs.
We query Snyk's API for each affected package and version,
matching advisories through a shared CVE identifier or an explicit
Snyk reference to GHSA.
Matching by CVE identifier recovers links missing from OSV,
which references Snyk for only \nOsvSnykRefs of the \nSnykLinked advisories
we link to Snyk.
We collect PoCs from both databases to avoid missing exploits available
in only one: \pctPocOnlySnyk of the linked advisories have a PoC only in
Snyk, while \pctPocOnlyGhsa have one only in GHSA.

We use GHSA as our primary advisory source. However,
to maintain compatibility with \vulcan~\cite{brito2023study} and
\secbench~\cite{bhuiyan2023secbench}, and ensure our collection is a superset of both, we also
include all vulnerabilities they contain, even when absent
from GHSA or excluded by our initial filters.
We match records across these datasets using their CVE, GHSA, or Snyk
identifiers~\cstep{3}~\cstep{4}.
This process adds \nAddedTotal vulnerabilities:
\nAddedWrongCwe are GHSA advisories with incorrect CWE labels,
\nAddedWithoutCwe are GHSA advisories without CWE labels, and
\nAddedSnykOnly are Snyk advisories with no GHSA counterpart.
Records describing the same vulnerability are merged into a single
entry, keyed on its CVE identifier or, when none was assigned, on its
GHSA or Snyk identifier. When sources disagree on advisory metadata,
GHSA takes precedence. As advisory CWE labels can be imprecise,
we manually validate the class assignments.
Of the resulting \nEntries entries, \nEntriesCve have a CVE,
\nEntriesGhsa a GHSA, and \nEntriesSnyk a Snyk identifier.

\begin{figure}[t]
  \centering
  \includegraphics[scale=0.75]{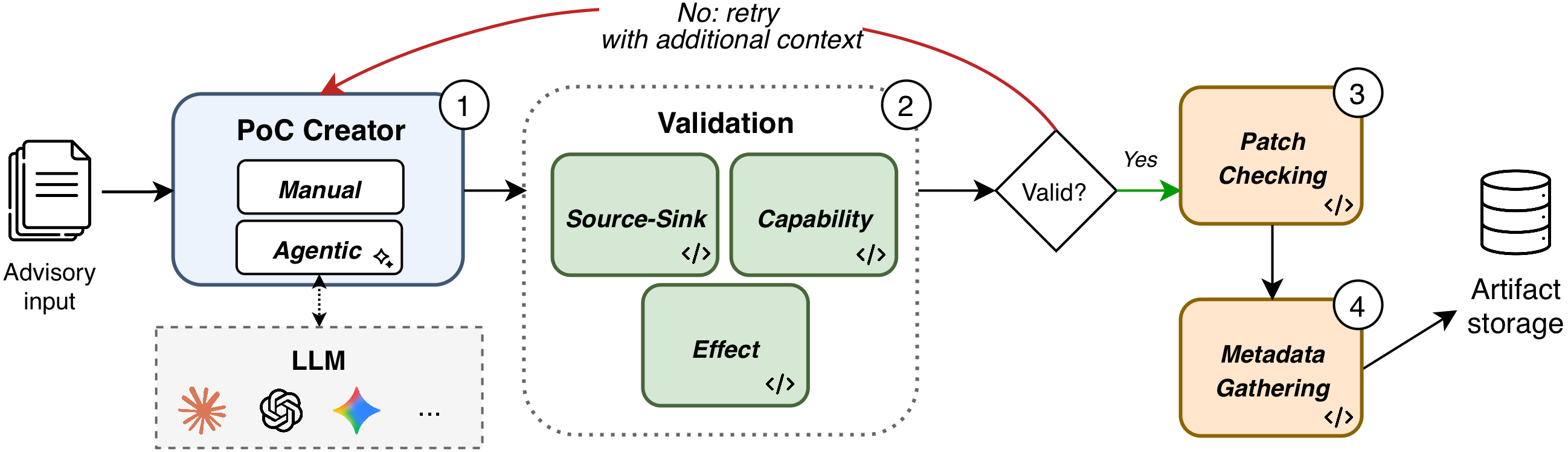}
  \caption{\system's PoC generation and validation pipeline.}
  \Description{The \system\ package analysis pipeline, with four stages:
  PoC creation (manual or LLM-based), validity analyses (source--sink,
  capability, and effect), patch checking, and metadata gathering. A
  decision node sends invalid exploits back to PoC creation with additional
  context.}
  \label{fig:pipeline}
  \vspace{-10pt}
\end{figure}

\subsection{Pipeline Overview}
\label{subsec:pipeline}

To produce valid exploits at scale, we built a pipeline with four stages
(Figure~\ref{fig:pipeline}). \emph{PoC Creator}
(\S\ref{sssec:poc-creator}) produces a candidate exploit, either through
an LLM agent or by hand, in a lifecycle format that separates harness
steps from the attack. \emph{Validation} then checks whether the exploit
is valid.
An accepted exploit proceeds to \emph{Patch Checking}
(\S\ref{sssec:patch}), which tests whether it still succeeds against
the patched version of the corresponding package, and to
\emph{Metadata Gathering} (\S\ref{sssec:metadata}), which records
the entry in the dataset.
Below, we describe phases 1, 3, and 4, and defer the more complex validation stage to \S\ref{subsec:validation}.

\subsubsection{PoC Creator}
\label{sssec:poc-creator}
PoC creation has two stages.
The first stage produces a free-form candidate exploit from its
advisory information: the textual description, the CWE class, the affected
package and vulnerable version, and any attached PoC. 
If no source provides a PoC, an LLM agent generates a free-form one from the advisory information.
We used
Claude Sonnet~4.6 as the agent, but our pipeline is model-agnostic: the agent
sits behind a fixed interface, so any sufficiently capable LLM can take its~place.
The second stage translates the free-form PoC, whether newly written or reused, into the
lifecycle format. The translation may only reorganize the exploit: it must
neither weaken the final check nor replace any part of the package with
placeholder code.

\subsubsection{Patch Checking}
\label{sssec:patch}
For every validated PoC, this stage identifies a candidate
patched version of the package and checks whether the PoC fails
against it.
When the advisory names a patched version, as it does for
\pctFixFromAdvisory of entries, we use that version. Otherwise, we
search the package's repository history for a candidate patching commit,
scanning commit messages for the keywords \texttt{fix},
\texttt{patch}, \texttt{security}, and \texttt{sanitize}, as well as
the advisory's~identifiers. 
We then run the lifecycle PoC against the patched code. If the PoC still
runs but its \texttt{validate} phase fails, it means the patch blocks the exploit and is thus labeled  \emph{valid}. 
If the exploit still succeeds, the patch is
kept but labeled \emph{invalid} to indicate that it does
not block the exploit. A valid label means that the patch blocks the tested
PoC, not that it closes every variant of the vulnerability.

\subsubsection{Metadata Gathering}
\label{sssec:metadata}
The last stage consolidates the pipeline results into a structured
metadata record, stored with the entry's lifecycle PoC, execution trace,
and patched version(s). The record contains the affected package and its
vulnerable version; the advisory identifiers (CVE, GHSA, and Snyk) and
CWE class; the source and sink locations found by the validation stage;
and, when available, the patched version together with its validity
label. The advisory identifiers allow each entry to be traced back to
the databases that reported it. The record also includes flags indicating
whether the vulnerability is present in \vulcan, \secbench, or both,
to facilitate comparisons across datasets in future studies.

\subsection{PoC Validation}
\label{subsec:validation}

We first introduce four concrete rules, (C1)--(C4),
that operationalize the validity conditions (i)--(iii) established in
\S\ref{ssec:motivation-poc-validation}:

\begin{figure}[t]
  \centering
  \begin{minipage}[c]{0.36\textwidth}
    \centering
    \includegraphics[width=0.9\linewidth]{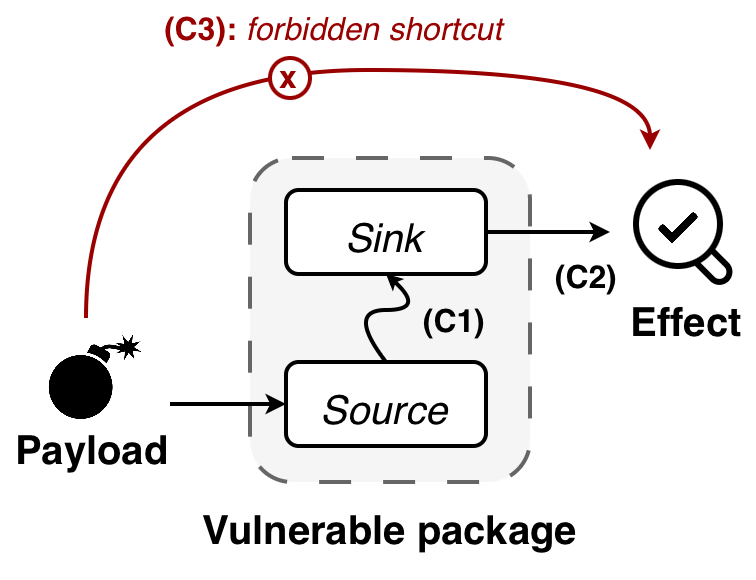}
  \end{minipage}\hspace{0.06\textwidth}%
  \begin{minipage}[c]{0.44\textwidth}
    \centering
    \input{Tables/capabilities.tex}
  \end{minipage}
  \caption{Exploit validity. \emph{Left:} a valid exploit reaches a sink from
  a source (C1) and produces an observable, checked effect (C2) through the
  vulnerable package, never by itself (C3). \emph{Right:} capabilities an
  exploit may not use (\capyes) for each vulnerability class: file-system
  read (FS-R) and write (FS-W), command execution (Cmd), runtime code evaluation
  (Eval), and prototype manipulation (PCM).}
  \label{fig:validity}
  \Description{Left: the exploit's payload enters the vulnerable package at a
  source, flows to a sink (C1), and produces an observable, checked effect
  (C2); a crossed-out arrow from the payload directly to the effect marks the
  forbidden case in which the exploit produces the effect without the package
  (C3). Right: a matrix of forbidden capabilities per CWE. CWE-22 forbids
  file read, file write, command execution, and runtime code evaluation;
  CWE-78 forbids command execution and runtime code evaluation; CWE-94
  forbids runtime code evaluation; CWE-1321 forbids runtime code evaluation
  and prototype manipulation.}
\end{figure}

\begin{itemize}[leftmargin=*]
  \item \emph{Source-to-sink flow (C1).} Attacker-controlled data must
  flow from a source (\S\ref{subsec:threat-model}) to a sink of the
  vulnerability class, e.g., a file-system operation for path traversal
  or \texttt{eval} for code injection, tying the exploit to the reported
  class and to concrete package locations. In
  Figure~\ref{fig:poc-format}, the source is the call to \texttt{setPath}
  (line~\ref{lst:ex-attack}) and the sink is the write inside
  \texttt{setPath} that sets \texttt{Object.prototype.polluted} to
  \texttt{yes}.

  \item \emph{Observable effect (C2).} The exploit must end by checking
  that the attack succeeded, e.g., that an injected command created a marker file or 
  that a fresh object carries the polluted property,
  as line~\ref{lst:ex-check} of Figure~\ref{fig:poc-format} does.

  \item \emph{No forbidden capabilities (C3).} To ensure the effect comes
  from the vulnerability rather than the harness, we introduce a novel
  capability-based constraint: the attack and check code may not use
  capabilities that directly produce the class's effect
  (Figure~\ref{fig:validity}, right). For instance, a prototype-pollution exploit may
  not assign to \texttt{Object.prototype}, and a path-traversal exploit may
  not read the target file directly. Code evaluation (\texttt{Eval}) is
  forbidden for all classes, since it can reproduce any effect. Setup and
  cleanup code may use these capabilities (\S\ref{subsec:threat-model}):
  in Figure~\ref{fig:poc-format}, cleanup deletes
  \texttt{Object.prototype.polluted}, but the pollution itself occurs
  inside \texttt{setPath}.

  \item \emph{Other constraints (C4).} The assertions made in the check code must fail when
  run before the attack, ruling out checks that always hold, and the
  attacker must use only the package's public~interface.
\end{itemize}

\noindent Each condition closes a gap the others leave open: a flow into a sink
(C1) does not show that the attack succeeds; an observed effect (C2) may
have been staged by the exploit itself, and not through the vulnerable package; 
and the absence of forbidden
capabilities (C3) does not show that anything happened. 
Figure~\ref{fig:validity} summarizes the rules. On the left, the payload
enters the package at a source and flows to a sink~(C1), which produces
the effect checked by the exploit~(C2); the crossed-out arrow is the
forbidden shortcut in which the exploit produces the effect without the
package~(C3). On the right, each row lists the capabilities that would
enable this shortcut for one vulnerability class. Path traversal (CWE-22) forbids the
most, since file reads, file writes, and commands all reach the target
file directly; command injection (CWE-78) forbids command execution;
prototype pollution (CWE-1321) forbids prototype manipulation; and Eval
is forbidden for every class, as it can reproduce any effect.
These restrictions apply to the attacker, not to the harness: setup may
prepare the environment, e.g., by creating a file the package will later
read, and validation inspects it afterwards. Such steps stand in for the
application around the package and for the analyst observing it; they are
not attacker actions.

\input{Figures/invalid-pocs}

Figure~\ref{fig:invalid-pocs} shows three PoCs rejected by the validation stage, each labeled with the rule it violates:
\begin{itemize}[leftmargin=*]
  \item \emph{Private access (a).} The PoC never goes through the
  package's public interface: it loads an internal module using a specified path 
  that does not belong to the package's public interface. 

  \item \emph{Forbidden call (b).} A command-injection PoC calls the
  package only as a decoy, without any effect:
  the file its \textcolor{phvalidate}{\texttt{validate}} phase will check is created by the PoC's own
  \texttt{execSync}, a call this vulnerability class forbids during
  \textcolor{phexploit}{\texttt{exploit}}.

  \item \emph{Vacuous check (c).} The PoC's assertion is a
  tautology that holds regardless of what happened, so \textcolor{phvalidate}{\texttt{validate}}
  already succeeds when run before the exploit.
\end{itemize}

\subsubsection{Basic analysis}
\label{sssec:basic-analysis}
We check the constraints above with a dynamic analysis built on top of
DynaJS~\cite{dynajs}, a shadow-execution framework for JavaScript that
improves on Jalangi~\cite{jalangi-fse13}. DynaJS instruments the PoC and
the package so that every relevant runtime event, such as a function call
or a property write, invokes an analysis hook, and propagates taint labels alongside each concrete value. On top of
these hooks we build two intermediate analyses, which all validity checks~use.

\runin{Trace Recording}The analysis records an \emph{ordered trace} of the
run: the functions entered, the calls made and their results, the
conditions evaluated, and the property writes, each tagged with its
source location and lifecycle phase. The trace is stored with each entry as evidence, so that a dataset consumer can
follow step by step how the attacker's input reached the vulnerable
operation.

\runin{Taint Propagation}The analysis dynamically tracks the taint status of runtime values,
pairing attacker-controlled values with taint labels. Values that the PoC passes into
the package during the \texttt{exploit} phase are marked as tainted, and any
operation on a tainted value taints its result, e.g., concatenating a
tainted string or reading a property of a tainted object. Taint labels
are recorded in the trace, so it shows which values the attacker
controlled at each step.

\subsubsection{Validity Checking}
\label{sssec:analysis-implementation}

This stage checks C1--C4 with two dynamic analyses: a
\emph{source--sink and effect analysis} for C1, C2, and C4, and a
\emph{capability analysis} for~C3.

\paragraph{Source--sink analysis (C1)}
We mark values as tainted when they enter the package through a public
entry point during \texttt{exploit}, treating that call as their
\emph{source}. For packages that receive attacker input through HTTP
requests, we also treat their HTTP handlers as sources and taint the
request objects they receive.
Sinks are defined specifically for each vulnerability class; for example, for path
traversal, they include Node.js file-system operations such as
\texttt{fs.readFile} and \texttt{fs.writeFile}, while for prototype pollution,
they are property writes targeting built-in objects.
Whenever a tainted value reaches a sink of the reported class, the
analysis records the source--sink flow. In the running example
(Figure~\ref{fig:poc-format}), the arguments of \texttt{setPath} are
tainted at entry. Taint propagates through the path segments as the
function traverses \texttt{target}, reaching the assignment of
\texttt{polluted} on \texttt{Object.prototype}.

To limit overhead on large packages, we initially instrument only the
target package, treating its dependencies as opaque calls
(Figure~\ref{fig:adaptive-scope}). If no source--sink flow of the reported
class is found, we instrument the dependencies imported by the executed
code and rerun the PoC. We repeat this process, extending instrumentation
one dependency layer at a time, until a flow is found. If no new imports
remain and no flow has been found, we instrument all dependencies.

\input{Figures/adaptive-scope}

\paragraph{Effect checking (C2)}
Reaching a sink does not establish that the attack produces the expected
effect. We therefore accept a flow only if the sink operation produces
that effect and the PoC's \texttt{validate} phase confirms it.
Concrete sink arguments identify what to check: an injected
\texttt{touch /tmp/pwned} names the file to create, so we check that it
is absent before the call and present afterwards. In the running
example, we detect that \texttt{polluted} appears on
\texttt{Object.prototype} and \texttt{validate} confirms its
presence on a fresh object.

\paragraph{Capability analysis (C3)}
We record the PoC's calls to sensitive APIs and attribute each to its
lifecycle phase, instrumenting the PoC and, when needed, its dependencies.
Each API maps to a capability: \texttt{fs.readFile}, for example,
provides file-system read access. Prototype manipulation is detected
through property writes to built-in prototypes.
We reject a PoC if it uses a capability forbidden for its vulnerability
class during \texttt{exploit} or \texttt{validate}
(Figure~\ref{fig:validity}). For path traversal, for example,
\texttt{setup} may create the target file, but \texttt{exploit} may
not read it directly.

\paragraph{Additional checks (C4)}
We rule out vacuous checks by running \texttt{validate} after
\texttt{setup} and before \texttt{exploit}, rejecting the PoC if it
already succeeds. In the running example, \texttt{validate} fails
before the exploit and succeeds afterwards.
We also check that the exploit uses only the public interface.
We load the package's public entry points, its main module and any
subpath declared in \texttt{exports}, and collect every function
reachable from the returned objects. As calls to valid entry points
return new objects, we repeat the search and extend this set.
Node.js module caching allows comparison by identity: every function
the PoC requires from the package must belong to the set.

%% file: Tables/capabilities.tex
\footnotesize
\setlength{\tabcolsep}{4pt}
\begin{tabular}{@{}l*{5}{c}@{}}
  \toprule
           & FS-R    & FS-W    & Cmd     & Eval    & PCM   \\
  \midrule
  CWE-22   & \capyes & \capyes & \capyes & \capyes & \capno  \\
  CWE-78   & \capno  & \capno  & \capyes & \capyes & \capno  \\
  CWE-94   & \capno  & \capno  & \capno  & \capyes & \capno  \\
  CWE-1321 & \capno  & \capno  & \capno  & \capyes & \capyes \\
  \bottomrule
\end{tabular}

%% file: Figures/invalid-pocs.tex
\newlength{\poccol}{\scriptsize\global\setlength{\poccol}{0.6em}}
\newcommand{\pocw}[1]{\dimexpr #1\poccol + 14pt\relax}
\lstdefinestyle{pocgrid}{style=pocplain,
  framesep=3pt, xleftmargin=3.6pt, framexleftmargin=0pt,
  xrightmargin=3.6pt, framexrightmargin=0pt}
\begin{figure}[t]
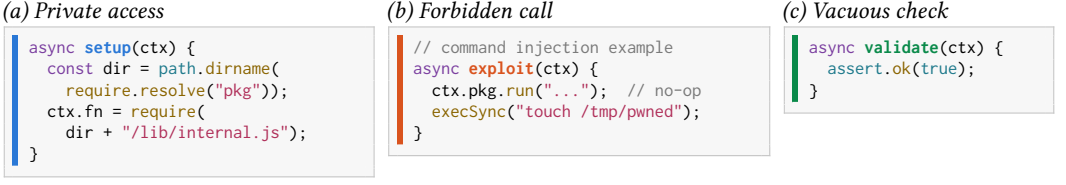

  \centering
  \begin{minipage}[t]{\pocw{30}}
{\small\textit{(a) Private access}}\par\vspace{2pt}
\begin{lstlisting}[style=pocgrid]
(*@\pbar{phsetup}@*)async setup(ctx) {
(*@\pbar{phsetup}@*)  const dir = path.dirname(
(*@\pbar{phsetup}@*)    require.resolve("pkg"));
(*@\pbar{phsetup}@*)  ctx.fn = require(
(*@\pbar{phsetup}@*)    dir + "/lib/internal.js");
(*@\pbar{phsetup}@*)}
\end{lstlisting}
  \end{minipage}\hfill
  \begin{minipage}[t]{\pocw{31}}
{\small\textit{(b) Forbidden call}}\par\vspace{2pt}
\begin{lstlisting}[style=pocgrid]
(*@\pbar{phexploit}@*)// command injection example
(*@\pbar{phexploit}@*)async exploit(ctx) {
(*@\pbar{phexploit}@*)  ctx.pkg.run("...");  // no-op
(*@\pbar{phexploit}@*)  execSync("touch /tmp/pwned");
(*@\pbar{phexploit}@*)}
\end{lstlisting}
  \end{minipage}\hfill
  \begin{minipage}[t]{\pocw{21}}
{\small\textit{(c) Vacuous check}}\par\vspace{2pt}
\begin{lstlisting}[style=pocgrid]
(*@\pbar{phvalidate}@*)async validate(ctx) {
(*@\pbar{phvalidate}@*)  assert.ok(true);
(*@\pbar{phvalidate}@*)}
\end{lstlisting}
  \end{minipage}
  \caption{Three invalid PoC snippets rejected by the validation stage, each labeled with the rule it violates.}
  \label{fig:invalid-pocs}
  \Description{Three short JavaScript listings side by side, in boxes sized
  to their code. (a) Private access: a setup phase that resolves the path of
  pkg and requires its internal file lib/internal.js directly. (b) Forbidden call, a command injection example: an exploit phase that calls
  ctx.pkg.run, which does nothing, and then runs execSync touch /tmp/pwned
  itself. (c) Vacuous check: a validate phase whose only statement
  is assert.ok(true).}
\end{figure}

%% file: Figures/adaptive-scope.tex
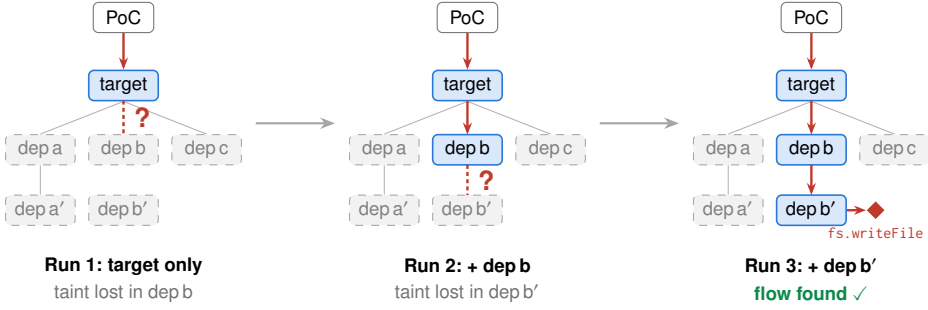
\begin{figure}[t]
\centering
\begin{tikzpicture}[
    font=\fontfamily{phv}\selectfont\scriptsize,
    pkg/.style={draw, rounded corners=2pt, minimum width=0.92cm,
                minimum height=0.40cm, inner sep=1pt},
    inst/.style={pkg, fill=scopeInst, draw=scopeInstB, line width=0.6pt},
    opq/.style={pkg, fill=black!5, draw=black!35, dashed, text=black!50},
    poc/.style={pkg, fill=white, draw=black!55, minimum width=0.8cm},
    tree/.style={draw=black!30, line width=0.4pt},
    taint/.style={-{Stealth[length=1.6mm]}, line width=0.9pt, draw=scopeTaint},
    lost/.style={line width=0.9pt, draw=scopeTaint, dash pattern=on 1.5pt off 1.2pt},
    sink/.style={diamond, fill=scopeTaint, inner sep=0pt, minimum size=2.4mm},
    qmark/.style={font=\fontfamily{phv}\selectfont\bfseries, text=scopeTaint},
    title/.style={font=\fontfamily{phv}\selectfont\scriptsize\bfseries, align=center},
    outcome/.style={font=\fontfamily{phv}\selectfont\scriptsize, align=center},
  ]
  \definecolor{scopeInst}{HTML}{DAE8FC}
  \definecolor{scopeInstB}{HTML}{2A78D6}
  \definecolor{scopeTaint}{HTML}{C0392B}
  \definecolor{scopeOK}{HTML}{0A8A4A}

  \newcommand{\scopepanel}[3]{%
    \begin{scope}[xshift=#1]
      \node[poc]  (poc) at (0, 0.9)    {PoC};
      \node[inst] (t)   at (0, 0)      {target};
      \node[opq]  (a)   at (-1.1,-0.85) {dep\,a};
      \node[#2]   (b)   at ( 0,  -0.85) {dep\,b};
      \node[opq]  (c)   at ( 1.1,-0.85) {dep\,c};
      \node[opq]  (a1)  at (-1.1,-1.65) {dep\,a$'$};
      \node[#3]   (b1)  at ( 0,  -1.65) {dep\,b$'$};
      \draw[tree] (t.south) -- (a.north);
      \draw[tree] (t.south) -- (c.north);
      \draw[tree] (a) -- (a1);
      \draw[taint] (poc) -- (t);
    \end{scope}}

  \scopepanel{0cm}{opq}{opq}
  \draw[lost] (0,-0.2) -- (0,-0.65);
  \node[qmark] at (0.25,-0.42) {?};
  \node[title]   at (0,-2.4) {Run 1: target only};
  \node[outcome, text=black!55] at (0,-2.75) {taint lost in dep\,b};

  \scopepanel{4.55cm}{inst}{opq}
  \begin{scope}[xshift=4.55cm]
    \draw[taint] (0,-0.2) -- (0,-0.65);
    \draw[lost]  (0,-1.05) -- (0,-1.45);
    \node[qmark] at (0.25,-1.25) {?};
  \end{scope}
  \node[title]   at (4.55,-2.4) {Run 2: + dep\,b};
  \node[outcome, text=black!55] at (4.55,-2.75) {taint lost in dep\,b$'$};

  \scopepanel{9.1cm}{inst}{inst}
  \begin{scope}[xshift=9.1cm]
    \draw[taint] (0,-0.2) -- (0,-0.65);
    \draw[taint] (0,-1.05) -- (0,-1.45);
    \node[sink] (s) at (0.85,-1.65) {};
    \draw[taint] (b1.east) -- (s.west);
    \node[font=\tiny, text=scopeTaint] at (0.85,-1.95) {\texttt{fs.writeFile}};
  \end{scope}
  \node[title]   at (9.1,-2.4) {Run 3: + dep\,b$'$};
  \node[outcome, text=scopeOK] at (9.1,-2.75) {\textbf{flow found} $\checkmark$};

  \draw[-{Stealth[length=2mm]}, line width=0.8pt, black!35] (1.75,-0.5) -- (2.8,-0.5);
  \draw[-{Stealth[length=2mm]}, line width=0.8pt, black!35] (6.3,-0.5) -- (7.35,-0.5);
\end{tikzpicture}
\vspace{-10pt}
\caption{Adaptive instrumentation: after each run without a flow, the
packages imported by code that executed are also instrumented (solid blue)
and the PoC runs again; dashed grey packages run as opaque calls. If no new
imports remain, all dependencies are instrumented.}
\Description{Three panels showing the same dependency tree: a PoC above a
target package with dependencies a, b and c, where a and b have their own
dependencies a-prime and b-prime. Run 1 instruments only the target and the
taint arrow into dependency b ends in a question mark. Run 2 also
instruments b, and the taint is lost in b-prime. Run 3 also instruments
b-prime, and the taint reaches a red sink labeled fs.writeFile, marked as
flow found. Dependencies a, a-prime and c stay opaque in every run.}
\label{fig:adaptive-scope}
\vspace{-10pt}
\end{figure}

%% file: Sections/Dataset.tex
\section{\system}
\label{sec:benchmark}

This section describes \system: its composition by vulnerability 
class (\S\ref{subsec:overview}), a manual check of its validity
analyses (\S\ref{subsec:sampling}), the advisories it excludes and the reasons
why (\S\ref{subsec:exclusions}), the structure of its entries 
(\S\ref{subsec:entry}), and how entries remain reproducible over 
time~(\S\ref{subsec:perenniality}).

\subsection{Overview}
\label{subsec:overview}

\system contains \nAllAdv advisories affecting \nAllPkgs packages, each
with at least one executable lifecycle PoC, for a total of \nAllPocs PoCs.\footnote{Advisories may affect multiple source-sink pairs and for those cases we try to collect one PoC per pair.}
Table~\ref{tab:overview} summarizes the dataset by vulnerability class.
Prototype pollution and path traversal together account for \dyn{two thirds}
of the advisories, reflecting their prevalence in the advisory databases.
At the time of writing, the majority of the advisories have PoCs that pass the validity analyses: \nAllCapOnly (\pctAllCapOnly) pass the capability analysis, \nAllFlow (\pctAllFlow) have a matching
source--sink flow, and \nAllCapab (\pctAllCapab) both. 
The remaining PoCs await manual inspection.
Code injection PoCs have the lowest pass rates in both analyses, partly because of their complex dynamic nature and the fact that many rely on capabilities with ambiguous intent, requiring manual inspection.
Patches are available
for \nAllPatches advisories, of which \nAllPatchValid (\pctAllPatchValid)
are confirmed to block the PoC. For the remaining, 94 patches do not block the PoC, and 41 fail installation or are no longer available. We consider them to be invalid patches.

Table~\ref{tab:comparison} compares \system with the two existing
benchmarks of Node.js vulnerabilities. 
 \system 
 combines source--sink annotations with executable PoCs. 
VulcaN~\cite{brito2023study}
provides neither PoCs nor patches.
\secbench~\cite{bhuiyan2023secbench} 
is missing source information and includes patches from the advisories without testing them.
\system substantially extends the coverage: %
\nAdvFromSecBench of its advisories appear in \secbench and
\nAdvFromVulcan in VulcaN (\nAdvBoth in both), and \nAdvNew appear in neither.

\input{Tables/dataset-overview}

\subsection{Manual Validation}
\label{subsec:sampling}

To assess the trustworthiness of the validity analyses, we manually
inspected approximately a tenth of each of the following groups of advisories: 
advisories that pass all the validity analyses (\nSampleValid out of \nAllCapab),
advisories without a matching source--sink flow (\nSampleNoFlow out of \nNoFlow), and 
advisories that have a matching flow but are rejected by the capability analysis (\nSampleInvalid out of \nNoCapability).
To keep each sample representative, we stratified
by vulnerability class and publication period (up to 2022 and from 2023), 
allocated the sample to each stratum in
proportion to its size, and drew advisories uniformly at random within
each stratum using a fixed seed. Since the allocation is proportional, results on
a sample estimate the whole group it was drawn from.

\input{Tables/benchmark-comparison}

All \nSampleValid PoCs that pass the analyses are valid. 
For the \nSampleNoFlow PoCs without a
flow, we observe that the missing flow reflects limits of the
analysis, or its inputs, rather than an invalid PoC: \nNoFlowTimeout of these timed out,
\nNoFlowNative reached a sink inside a native binary that our analysis cannot
instrument (we manually confirmed these), in \nNoFlowCrash cases the analysis crashed, 
\nNoFlowCwe advisories had a wrong CWE label, and \nNoFlowSink used a sink missing from our catalog; 
we have corrected the last two. Of the \nSampleInvalid PoCs rejected by the
capability analysis, \nInvalidTrue were indeed invalid. The remaining one was
a command-injection PoC that calls \texttt{RegExp.prototype.exec}, which the
analysis misclassified as the \texttt{exec} function of
\texttt{child\_process}. We patched the capability analysis to rule out such
name clashes, after which this entry was validated.

\subsection{Exclusion Criteria}
\label{subsec:exclusions}

We exclude an advisory from our dataset (\nExcl out of \nEntries) whenever its exploit cannot be reproduced, or when the
package alone cannot demonstrate the vulnerability:

\begin{itemize}[leftmargin=2.6em, labelwidth=2.1em, labelsep=0.5em, align=left, itemsep=3pt]
  \item[\textbf{(E1)}] \emph{No longer on npm} (\nExclUnpublished).
  The vulnerable version, or the whole package, has been removed from the
  registry at the time of our data collection, so it cannot be installed.

  \item[\textbf{(E2)}] \emph{Installation fails} (\nExclInstall). The
  vulnerable version is still published but no longer installs.

  \item[\textbf{(E3)}] \emph{Requires a user interface} (\nExclGui).
  Triggering the vulnerability requires interacting with a user interface, 
  which our headless environment cannot reproduce.

  \item[\textbf{(E4)}] \emph{Insufficient on its own} (\nExclGenerated).
  The package produces something unsafe, but the harm only occurs when
  another component consumes it.
\end{itemize}

We illustrate \textbf{E4} with
\texttt{shell-quote} (CVE-2016-10541), which turns a list of arguments
into a string to be passed to a shell. Incorrect escaping allows
an attacker-controlled command to be included in the returned string.
While the flaw is critical, \texttt{shell-quote} never executes that string:
the command runs only when an application passes it to a function such
as \texttt{child\_process.exec}. A PoC demonstrating the vulnerability
must therefore execute the string, using the capability that
condition C3 forbids. No source--sink flow for command execution exists
within the package. 
We exclude such
vulnerabilities because they only manifest 
with a consuming application.

\subsection{Anatomy of an Entry}
\label{subsec:entry}

\system is organized by package version: each directory contains one
vulnerable version of a package, and its \texttt{vulnerability.json}
contains one record per advisory affecting that version. Each record
defines an \emph{entry}, linking the advisory to its PoC, trace, and
patched version, when available. An advisory that names several entry
points, such as two exported functions that each reach a sink, has one
record per entry point: all share the advisory identifiers, but each has
its own PoC (\texttt{poc01/}, \texttt{poc02/}, \dots), the source and sink
locations that PoC reaches, and its own patch check. \nAdvMultiEntry
advisories have more than one record. Figure~\ref{fig:entry} shows the directory for
\texttt{ini}~1.3.5, which contains a single entry for the prototype
pollution vulnerability CVE-2020-7788. An entry contains the following~artifacts:

\begin{itemize}[leftmargin=*, itemsep=3pt]
  \item \emph{Metadata} (the record in \texttt{vulnerability.json}):
  the advisory identifiers and CWE, the source and sink locations
  identified by the validity analyses, and information about the patch version,
  when available.
  \item \emph{Exploit} (\texttt{poc01/}): the lifecycle PoC, with a
  \texttt{package.json} that pins the vulnerable version and any necessary dependencies.
  \item \emph{Trace} (\texttt{coverage.json}, alongside the PoC):
  the ordered execution trace of the PoC, showing how attacker input
  travels from source to sink.
  \item \emph{Patch} (\texttt{patch01/}, when available):
  a \texttt{package.json} that pins the patched version and the diff
  of the patch commit.
\end{itemize}

\input{Figures/entry-anatomy}

Each entry supports several uses. The recorded source and sink locations
provide a reference for checking whether detection tools identify the
vulnerable code correctly. The PoCs can be used to test whether a
mitigation blocks the exploits. 
The patched versions and recorded patches can be used to evaluate
both detection and automatic repair tools: detection tools can be
checked for reports of vulnerabilities that have been patched, while
generated repairs can be tested against the PoCs and compared with
the recorded patches.
Finally, the trace supports
detailed studies of how each vulnerability unfolds, including the
complexity measures in \S\ref{sec:metrics}.

\subsection{Perenniality}
\label{subsec:perenniality}

PoCs depend on the availability of the packages they install. If a required
version of a target package or dependency is removed from npm, the PoC
may no longer run. In fact, for \nExclUnpublished advisories the
vulnerable version was already unavailable during collection (E1).
To keep entries runnable, \system uses a self-hosted npm registry based
on Verdaccio~\cite{verdaccio}. The registry preserves the vulnerable and
patched versions of each target package, along with all transitive
dependencies used by the PoCs.

Packages are added to the registry when new advisories are processed
by the pipeline in \S\ref{sec:methodology}. During the initial installation,
Verdaccio downloads the requested package tarballs from npm and caches
them locally. npm also generates a \texttt{package-lock.json} file
recording the exact versions installed, including transitive dependencies.
To reproduce the original test results, dependencies for confirmed PoCs should 
be installed using the corresponding lockfile and the self-hosted registry. 
This ensures that the exact package versions used during testing are installed, 
even if they are no longer available on npm.

%% file: Tables/dataset-overview.tex
\begin{table}[t]
  \caption{Contents of \system per vulnerability class, counted in advisories.}
  \label{tab:overview}
  \small
  \setlength{\tabcolsep}{3.5pt}
  \begin{tabular*}{\textwidth}{@{\extracolsep{\fill}}l rr rrr rrr@{}}
    \toprule
    & & & \multicolumn{3}{c}{Validated} & \multicolumn{3}{c}{Patches} \\
    \cmidrule(lr){4-6} \cmidrule(l){7-9}
    Class & Advisories & Pkgs. & Capabilities & Flow & Cap. $\cap$ Flow & Total & Valid & Invalid \\
    \midrule
    Prototype pollution & \nPPAdv & \nPPPkgs
      & \nPPCapOnly~{\color{black!45}\scriptsize(\pctPPCapOnly)} & \nPPFlow~{\color{black!45}\scriptsize(\pctPPFlow)} & \nPPCapab~{\color{black!45}\scriptsize(\pctPPCapab)}
      & \nPPPatches & \nPPPatchValid~{\color{black!45}\scriptsize(\pctPPPatchValid)} & \nPPPatchInvalid~{\color{black!45}\scriptsize(\pctPPPatchInvalid)} \\
    Path traversal & \nPTAdv & \nPTPkgs
      & \nPTCapOnly~{\color{black!45}\scriptsize(\pctPTCapOnly)} & \nPTFlow~{\color{black!45}\scriptsize(\pctPTFlow)} & \nPTCapab~{\color{black!45}\scriptsize(\pctPTCapab)}
      & \nPTPatches & \nPTPatchValid~{\color{black!45}\scriptsize(\pctPTPatchValid)} & \nPTPatchInvalid~{\color{black!45}\scriptsize(\pctPTPatchInvalid)} \\
    Command injection & \nCmdIAdv & \nCmdIPkgs
      & \nCmdICapOnly~{\color{black!45}\scriptsize(\pctCmdICapOnly)} & \nCmdIFlow~{\color{black!45}\scriptsize(\pctCmdIFlow)} & \nCmdICapab~{\color{black!45}\scriptsize(\pctCmdICapab)}
      & \nCmdIPatches & \nCmdIPatchValid~{\color{black!45}\scriptsize(\pctCmdIPatchValid)} & \nCmdIPatchInvalid~{\color{black!45}\scriptsize(\pctCmdIPatchInvalid)} \\
    Code injection & \nCodeIAdv & \nCodeIPkgs
      & \nCodeICapOnly~{\color{black!45}\scriptsize(\pctCodeICapOnly)} & \nCodeIFlow~{\color{black!45}\scriptsize(\pctCodeIFlow)} & \nCodeICapab~{\color{black!45}\scriptsize(\pctCodeICapab)}
      & \nCodeIPatches & \nCodeIPatchValid~{\color{black!45}\scriptsize(\pctCodeIPatchValid)} & \nCodeIPatchInvalid~{\color{black!45}\scriptsize(\pctCodeIPatchInvalid)} \\
    \midrule
    \textbf{Total} & \textbf{\nAllAdv} & \nAllPkgs
      & \nAllCapOnly~{\color{black!45}\scriptsize(\pctAllCapOnly)} & \nAllFlow~{\color{black!45}\scriptsize(\pctAllFlow)} & \nAllCapab~{\color{black!45}\scriptsize(\pctAllCapab)}
      & \nAllPatches & \nAllPatchValid~{\color{black!45}\scriptsize(\pctAllPatchValid)} & \nAllPatchInvalid~{\color{black!45}\scriptsize(\pctAllPatchInvalid)} \\
    \bottomrule
  \end{tabular*}
  \par\smallskip\raggedright\footnotesize
\end{table}

%% file: Tables/benchmark-comparison.tex
\newcommand{\yes}{$\checkmark$}
\newcommand{\no}{{\color{black!40}$\times$}}
\begin{table}[t]
  \caption{\system compared with existing Node.js vulnerability benchmarks.}
  \label{tab:comparison}
  \small
  \begin{tabular*}{\textwidth}{@{\extracolsep{\fill}}l r r l c r r c@{}}
    \toprule
    Benchmark & Vulns.$^a$ & PoCs & PoC validation & Source & Sink & Patches & Live \\
    \midrule
    VulcaN~\cite{brito2023study} & 313 {\color{black!45}\scriptsize(957)} & \no & -- & \yes$^b$ & \yes$^b$ & \no & \no \\
    \secbench~\cite{bhuiyan2023secbench} & 502 {\color{black!45}\scriptsize(600)} & 600 & Manual & \no & 567 & 285$^c$ & \no \\
    \midrule
    \textbf{\system} & \nAllAdv & \nAllPocs & Automated + Manual & \nAllFlow & \nAllFlow & \nAllPatchValid & \yes \\
    \bottomrule
  \end{tabular*}
  \par\smallskip\raggedright\footnotesize
  $^a$\,In the four classes of \system; in parentheses, across all classes the benchmark covers.
  $^b$\,Annotated manually; not counted per class.
  $^c$\,Patched versions reported by advisories, not checked against the exploit. 
\end{table}

%% file: Figures/entry-anatomy.tex
\definecolor{tcode}{HTML}{C0561C}
\definecolor{blue}{HTML}{007FFF}
\definecolor{green}{HTML}{177245}
\newcommand{\lbl}[1]{{\color{black!50}#1}}
\begin{figure}[t]
  \centering
  \begin{tikzpicture}[baseline=(current bounding box.center), x=1pt,y=1pt, fn/.style={font=\ttfamily\scriptsize, anchor=west, inner sep=0pt},
      dirn/.style={fn, font=\ttfamily\scriptsize\bfseries}, br/.style={draw=black!35, line width=0.5pt}]
    \draw[fill=black!3, draw=black!22, line width=0.4pt] (0,0) rectangle (112,-80.7);
    \node[dirn] at (6.0,-7.5) {ini-1.3.5/};
    \node[fn] at (21.0,-15.8) {vulnerability.json};
    \node[dirn] at (21.0,-24.1) {poc01/};
    \node[fn] at (32.0,-32.4) {package.json};
    \node[fn] at (32.0,-40.7) {poc01.js};
    \node[fn] at (32.0,-49.0) {coverage.json};
    \node[dirn] at (21.0,-57.3) {patch01/};
    \node[fn] at (32.0,-65.6) {package.json};
    \node[fn] at (32.0,-73.9) {patch01.patch};
    \draw[br] (9.0,-15.8) -- (19.0,-15.8);
    \draw[br] (9.0,-24.1) -- (19.0,-24.1);
    \draw[br] (24.0,-32.4) -- (30.0,-32.4);
    \draw[br] (24.0,-40.7) -- (30.0,-40.7);
    \draw[br] (24.0,-49.0) -- (30.0,-49.0);
    \draw[br] (9.0,-57.3) -- (19.0,-57.3);
    \draw[br] (24.0,-65.6) -- (30.0,-65.6);
    \draw[br] (24.0,-73.9) -- (30.0,-73.9);
    \draw[br] (9.0,-10.7) -- (9.0,-57.3);
    \draw[br] (24.0,-27.3) -- (24.0,-49.0);
    \draw[br] (24.0,-60.5) -- (24.0,-73.9);
  \end{tikzpicture}\hspace{14pt}%
  \begin{tikzpicture}[baseline=(current bounding box.center), x=1pt,y=1pt]
    \node[fill=black!3, draw=black!22, line width=0.4pt,
      minimum height=80.7pt, inner xsep=8pt, inner ysep=0pt, font=\ttfamily\scriptsize] {%
      \renewcommand{\arraystretch}{1.12}%
      \begin{tabular}{@{}l@{\hspace{10pt}}l@{}}
        \multicolumn{2}{@{}l}{\color{blue}\ttfamily vulnerability.json}\\[1pt]
        \lbl{Advisory} & CVE-2020-7788\enspace{\color{black!40}$\cdot$}\enspace GHSA-qqgx-2p2h-9c37 \\
        \lbl{Class}    & CWE-1321, prototype pollution \\
        \lbl{Source}   & \texttt{ini.js:69}\enspace\enspace {\color{tcode}\texttt{`function decode (str) \{'}} \\
        \lbl{Sink}     & \texttt{ini.js:109}\enspace {\color{tcode}\texttt{`p[key] = value'}} \\
        \lbl{Exploit}  & \texttt{poc01/poc01.js}\enspace{\color{black!40}$\cdot$}\enspace runs on \texttt{ini@1.3.5} \\
        \lbl{Patch}      & \texttt{1.3.6}\enspace{\color{black!40}$\cdot$}\enspace \color{green}blocks the PoC\enspace{\color{phvalidate}$\checkmark$}
      \end{tabular}};
  \end{tikzpicture}
  \caption{The directory of \texttt{ini@}1.3.5 in \system (left) and its single entry (right).}
  \label{fig:entry}
  \Description{Left: the directory for ini version 1.3.5, containing
  vulnerability.json, a poc01 folder with package.json, poc01.js and
  coverage.json, and a patch01 folder with package.json and patch01.patch.
  Right: the directory's single entry, the record in vulnerability.json for
  one advisory: CVE-2020-7788 and GHSA-qqgx-2p2h-9c37, CWE-1321 prototype
  pollution, source ini.js line 69 (function decode), sink ini.js line 109
  (p[key] = value), exploit poc01/poc01.js running on ini 1.3.5, and fix
  1.3.6, which blocks the PoC.}
\end{figure}
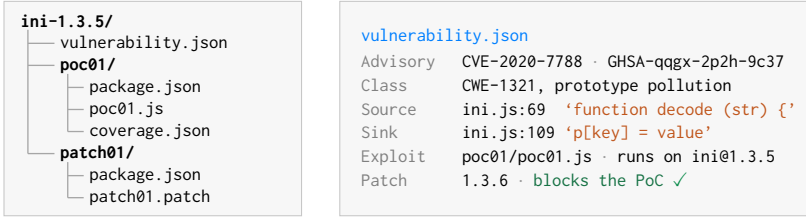

%% file: Sections/Measurements.tex
\section{Vulnerability Trends}
\label{sec:metrics}
We examine longitudinal trends in Node.js taint-style vulnerabilities, including their complexity (\S\ref{subsec:trend-complexity}), time-to-patch (\S\ref{subsec:trend-fixes}), repeated occurrence of the same vulnerability class within a package (\S\ref{subsec:trend-reintroduction}), and patches that fail to block their exploit (\S\ref{subsec:trend-zerodays}), using data collected by \system.

\subsection{Dataset and Setup}
We analyze the \nAllCapOnly advisories whose PoC succeeds on the vulnerable version and passes the capability
analysis; \nCapOnlyUndated cases do not have a publication
date and therefore are omitted from the per-year figures. We keep those without
a matching source--sink flow, since a missing flow reflects a limit of our
flow analysis rather than of the PoC (\S\ref{subsec:sampling}).
Most figures report the median and interquartile range for each advisory publication year. As only \nTrendAdvPreEighteen advisories were published before 2018, the sample size is too small for year-by-year analysis; we therefore group them with 2018 advisories and label the resulting category as $\leq$'18 in the figures. Data for 2026 are currently available only through July.

We take an advisory's patch to be the code that changed between its
vulnerable version and the patched version recorded in its entry
(\S\ref{sssec:patch}). We filter out non-JavaScript files, tests, and
documentation, but a release may still bundle unrelated changes, hence patch
sizes can overstate the patch itself.

Advisories in recent years are increasingly concentrated in a few
packages. The five packages with the most advisories account for
\pctTopFiveTwentyOne of the advisories published in 2021 and
\pctTopFiveTwentySix in 2026. The AI agent
\textit{OpenClaw}~\cite{openclaw2026} alone accounts for
\nOpenclawTwentySix advisories in 2026. 
To assess whether packages with many advisories disproportionately influence the results, we repeat each analysis, when applicable, using a single observation per package, thereby assigning equal weight to all packages. We highlight any differences from the results obtained using individual advisories.

\subsection{Vulnerability Complexity}
\label{subsec:trend-complexity}

\begin{figure}[t]
  \centering
  \includegraphics[width=0.9\linewidth]{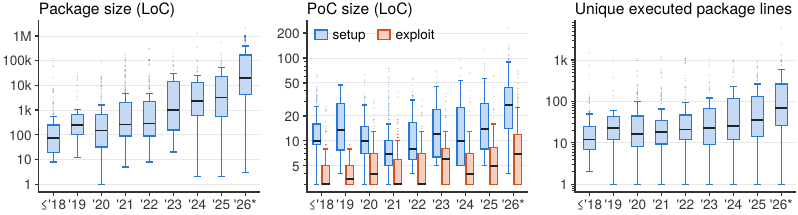}
  \caption{Complexity of \system's vulnerabilities by advisory publication
  year, one observation per advisory, on a logarithmic scale. Boxes span the
  interquartile range, black lines mark the median, whiskers extend to the
  most extreme value within 1.5 times the interquartile range, and dots are
  outliers.}
  \label{fig:trend-complexity}
  \Description{Three panels of yearly box plots over advisory publication
  years from 2018 (including earlier years) to 2026, on logarithmic axes.
  The median package size grows from under 100 to more than 20,000 source
  lines. PoC size is shown as paired boxes for the setup and exploit phases;
  the median setup grows from about 10 to about 30 lines, the median exploit
  from about 3 to about 7. The median number of unique executed package lines
  grows from about 12 to about 65.}
\end{figure}

Figure~\ref{fig:trend-complexity} shows trends in package size, the size of the
PoC's \texttt{setup} and \texttt{exploit} phases, and the number of
distinct package lines executed by the exploit.  All four metrics increase over time. 
The clearest trend is an increase in vulnerable package size. The median
package size grew from \medPkgSlocEighteen LoC in 2018 and
earlier and roughly \medPkgSlocEarlyAdv in 2021--2022 to
\medPkgSlocTwentyFourAdv in 2024 and \medPkgSlocTwentySix in~2026.
When packages with multiple advisories are counted only once, the median package size is consistently smaller than when every advisory is counted, by \pctPkgGapTwentyOne in 2021, \pctPkgGapTwentyTwo in 2022, and \pctPkgGapTwentyFour in 2024. The effect becomes much more pronounced in 2026, where the median decreases by \pctNoOpenclawDropTwentySix (from \medPkgSlocTwentySix to \medPkgSlocTwentySixNoOpenclaw) when \texttt{openclaw} is excluded and by \pctPerPkgDropTwentySix (to \medPkgSlocLate) when each package is counted only once.
This suggests that much of the increase observed in 2026 is driven by a small number of large packages with many advisories. Nevertheless, the long-term trend remains clear: advisories in our dataset increasingly affect large applications and frameworks rather than small utility libraries.

PoC sizes fluctuate before increasing steadily from 2024 to 2026. 
The attack itself remains short:
the median \texttt{exploit} phase grew from \medExploitEarly lines
before 2023 to \medExploitLate in 2026. The \texttt{setup} phase,
however, grew from \medSetupEarly lines in 2021--2022 to
\medSetupLate in 2026, or \medSetupTwentySixPkg when each package
is counted only once. This suggests that reaching these vulnerabilities
requires more setup, such as starting a server, configuring a database,
or preparing the environment. The exploits also execute more package
code: a median of \medExecLate distinct lines in 2026, compared with
\medExecEarly in 2021--2022. This trend persists with one observation
per package (\medExecLatePkg in 2026).

Together, these trends show that recent PoCs target larger packages,
require more setup, and execute more package code. These cases are
largely missing from earlier benchmarks, whose coverage ends in
2022~\cite{brito2023study,bhuiyan2023secbench}. Evaluation on those
benchmarks therefore gives limited evidence of how tools perform
on these more recent cases.

\takeaway{Recent advisories increasingly affect larger packages,
and their PoCs require more setup and execute more package code.}

\subsection{Patch sizes and timing}
\label{subsec:trend-fixes}

\begin{figure}[t]
  \centering
  \begin{minipage}[t]{0.49\linewidth}
    \centering
    \includegraphics[width=\linewidth]{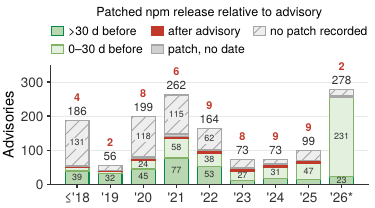}
    \caption{Release of the first patched npm version relative to the
    advisory's publication. Numbers are advisory counts; red numbers above
    each bar count the patches released after the advisory.}
    \label{fig:trend-fix-timing}
    \Description{A stacked bar chart of advisory counts per publication
    year, from 2018 (including earlier years) to 2026, split into advisories
    whose patched npm version was released more than 30 days before the
    advisory, 0 to 30 days before, after the advisory, with a patch but no
    release date, or with no fix recorded. Advisories with no fix recorded
    fall from 139 of 197 up to 2018 to 19 of 296 in 2026, while fixes
    released 0 to 30 days before the advisory rise to 247 in 2026. Between 2
    and 10 fixes per year are released after the advisory.}
  \end{minipage}\hfill
  \begin{minipage}[t]{0.49\linewidth}
    \centering
    \includegraphics[width=\linewidth]{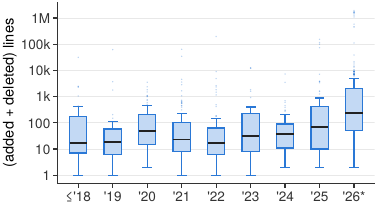}
    \caption{Size of the fixing patch (added plus deleted lines), one
    observation per advisory with a patch; box plots as in
    Figure~\ref{fig:trend-complexity}.}
    \label{fig:trend-churn}
    \Description{Yearly box plots, from 2018 (including earlier years) to
    2026, of the number of added plus deleted lines in the fixing patch, on a
    logarithmic axis. The median stays between about 17 and 48 lines until
    2024, then rises to about 70 in 2025 and about 260 in 2026.}
  \end{minipage}
\end{figure}

Figure~\ref{fig:trend-fix-timing} compares the release date of the first
patch version with the advisory publication date. Two trends stand
out. First, advisories without a recorded patch have become less common:
their share fell from \pctNoFixEighteen up to 2018 and
\pctNoFixTwentyOne in 2021 to \pctNoFixTwentySix in 2026.
Second, patches are now released closer to advisory publication.
Through 2022, advisories were typically published months after the patch:
the median gap was \medLagEighteen days up to 2018 and
\medLagTwentyOne days in 2021. Only \pctShortBeforeTwentyOne of the
2021 patches were released within the 30 days before the advisory.
Since 2023, the median gap has narrowed to \medLagRecent days;
in 2026, \pctShortBeforeTwentySix of the patches were released within
that 30-day window. 
The shorter gaps in recent years
are consistent with disclosure coordinated with maintainers.
Having patches released after advisory publication remains uncommon:
they account for at most \pctAfterMax of patches with known release
dates in any year (red numbers in Figure~\ref{fig:trend-fix-timing}).

Patches have also grown bigger (Figure~\ref{fig:trend-churn}).
The median number of changed lines (additions plus deletions) rose
from \medChurnEarly up to 2022 to \medChurnTwentyFive in 2025
and \medChurnTwentySix in 2026. Packages with many advisories account
for part of the 2026 increase. With one observation per package,
the median falls to \medChurnTwentySixPkg lines, but remains several
times the earlier value.

The shift towards larger packages may help explain this increase.
AI coding assistants are another possible factor: their adoption
has been linked to a sustained increase in code complexity in
open-source projects~\cite{he2026speed}. To examine their involvement
in patches, we checked the linked commits and pull requests for explicit
AI attribution, such as an agent credited as a co-author or mentioned
in the description. We found no such attribution before 2025,
\nAiTwentyFive cases in 2025, and \nAiTwentySix in 2026, among the
\nPatchedTwentySix advisories of that year whose patch we could measure. These counts
capture only disclosed use and may underestimate AI involvement.
In 2025--2026, patches with documented AI involvement changed a median
of \medChurnAi lines, compared with \medChurnNoAi for other patches.
However, there are too few such cases to establish a link between
AI involvement and patch size.

\takeaway{Advisories without a recorded patch have become less common,
falling from \pctNoFixEighteen up to 2018 to \pctNoFixTwentySix
in 2026. Recent advisories are typically published days after a patch
is released, rather than months later. Fixing patches have also grown in size.}

\subsection{Repeated Occurrences of Vulnerabilities within a Package}
\label{subsec:trend-reintroduction}

We observe that some packages appear repeatedly in advisories, including multiple advisories involving the same vulnerability class.
We define an advisory as a \emph{recurrence} if it affects the same package and vulnerability class (CWE) as a prior advisory and impacts a version released after the earlier vulnerability had been patched.
We require that patch to block the earlier PoC. This captures repeated
vulnerability classes, without implying that the same bug was
reintroduced: the advisories may concern different code or functionality.

\begin{figure}[t]
  \centering
  \begin{minipage}[t]{0.49\linewidth}
  \centering
  \includegraphics[width=\linewidth]{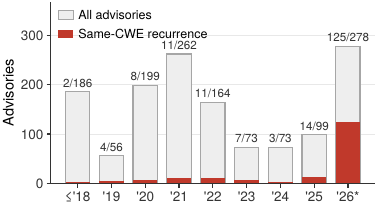}
  \caption{Recurrences: advisories for a vulnerability class that an
  earlier, patched advisory already reported in the same package.}
  \label{fig:trend-recurrence}
  \Description{A bar chart per publication year, from 2018 (including earlier years) to 2026, of all
  analyzed advisories, with an overlaid bar for recurrences of the same CWE in the same package,
  labeled with recurrences over all advisories. The
  overlaid bars are small until 2025 (between 3 and 14 advisories per year)
  and jump to 139 of 296 advisories in 2026.}
  \end{minipage}\hfill
  \begin{minipage}[t]{0.49\linewidth}
  \centering
  \includegraphics[width=\linewidth]{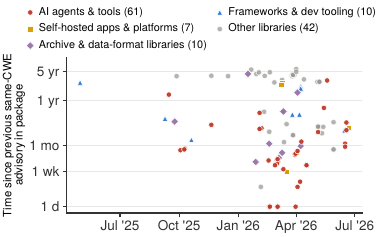}
  \vspace{-10pt}
  \caption{Time since the previous same-CWE advisory in the same package,
  for recurrences since 2025.}
  \label{fig:trend-recurrence-gap}
  \Description{A scatter plot of recurrences from mid-2025 to mid-2026. The
  horizontal axis is the publication date; the vertical axis, on a
  logarithmic scale from one day to five years, is the time since the
  previous advisory of the same CWE in the same package. Points are
  colored by kind of package: AI agents and tools (72), self-hosted apps
  and platforms (9), archive and data-format libraries (10), frameworks and
  developer tooling (11), and other libraries (42). AI agent points
  concentrate between one day and one month, mostly from February to June
  2026; points for other libraries concentrate between one and five years,
  mostly from February to May 2026.}
  \end{minipage}
\end{figure}

Figure~\ref{fig:trend-recurrence} shows that recurrences accounted for
\pctRecurEarlyRange of each year's advisories through 2024 and
\pctRecurTwentyFive in 2025. In 2026, this share rose to
\pctRecurTwentySix (\nRecurTwentySix of \nTrendAdvTwentySix).
About 40\% came from \texttt{openclaw}, which has
\nRecurOpenclaw recurrences among its \nOpenclawTwentySix advisories.
The increase extends beyond this package. Excluding \texttt{openclaw},
\pctRecurTwentySixNoOpenclaw of the 2026 advisories are recurrences,
and \pctPkgRecurTwentySix of the affected packages have at least one,
compared with at most \pctPkgRecurEarly in earlier years.
Advisories also cluster in well-established libraries: \texttt{handlebars} and \texttt{vm2} each received five advisories within a four-day period. Such clusters are consistent with audits that uncover multiple related vulnerabilities at once.

Figure~\ref{fig:trend-recurrence-gap} measures the time between each
of the \nGapRecur recurrences published since 2025 and the previous
advisory for the same package and CWE, omitting the \nGapSameDay whose previous advisory was published on the same day or later. Two patterns stand out.
The \nGapAI recurrences in AI agents and tools followed the previous advisory
by a median of \medGapAI days, and \pctGapAIMonth of them within a
month. The \nGapOther recurrences in other libraries span
\nGapOtherPkgs packages, including \texttt{axios}, \texttt{handlebars},
\texttt{vm2}, and \texttt{lodash}. Their median gap was
\medGapOtherYears years, and \pctGapOtherYear of them came more than a
year after it. Most of these longer-gap
recurrences (\nGapOtherLongSpring of \nGapOtherLong) were published
between February and May 2026, following earlier advisories published
in 2019--2023.

A surge in using AI-assisted vulnerability detection and software development may help explain these patterns.
However, \system lacks necessary data to establish either as a cause. 
Coincidentally, AI models have been
used to audit open-source code at scale~\cite{anthropic2026firefox}.
One campaign reported thousands of previously unknown flaws in
JavaScript and TypeScript projects, concentrated in code injection
and prototype pollution~\cite{unit42_2026burst}. Such audits could
explain why several advisories appear together in libraries whose
previous advisories were published years earlier.
AI-assisted development offers another possible explanation.
Studies have found that coding agents can produce insecure code
that is functionally correct~\cite{pearce2025asleep,zhao2025vibe,perry2023users,chen2026securevibebench}.
If used in these packages, they could contribute to repeated
vulnerabilities of the same class. However, short gaps between
advisories could also reflect closer scrutiny of popular packages.

\takeaway{Almost half of the advisories published in 2026 concern
a vulnerability class previously patched in the same package, compared
with at most 10\% in each year through 2024. The gap between these
advisories is typically days in AI agents and tools, but years
in established libraries.}

\subsection{Incomplete Patches}
\label{subsec:trend-zerodays}

Patch checking (\S\ref{sssec:patch}) found \nPatchNoBlock advisories
whose PoC still succeeds on the version the advisory reports as patched.
Such a failure can have benign causes, such as an advisory that records
the wrong version, but it can also mean that the patch is incomplete. We
therefore re-ran these PoCs on the latest version of each package
published on npm. For two packages, the PoC still succeeded, even though
every advisory for them lists the latest version as not vulnerable.
At the time of our report, these latest, still-vulnerable releases each had
more than 980k and 500k weekly downloads, respectively.
Both vulnerabilities are described in
advisories, but the fact that their patches do not work is not: users who
upgrade to the version listed as safe remain exposed, so for anyone who
relies on advisory data these behave like zero-days. We are reporting both
cases to their maintainers and to the GitHub Advisory Database through
coordinated disclosure, and withhold the package names and PoCs until the
advisories are corrected.

%% file: Sections/Discussion.tex
\section{Discussion}
\label{sec:discussion}

\paragraph{Limitations.}
Our collection targets four taint-style vulnerability classes and draws
predominantly on reviewed GHSA records. We use the advisory's CWE labels to identify
candidate vulnerabilities, so missing or incorrect labels may prevent
relevant cases from being found. Recovering records from other sources, such as
\vulcan~\cite{brito2023study} and \secbench~\cite{bhuiyan2023secbench},
reduces this risk but does not remove it.
\system\ also excludes vulnerabilities whose harmful operation occurs only
in a consuming application, as with the \texttt{shell-quote} example in
\S\ref{subsec:exclusions}, and those that require the exploit to be delivered through a user interface.
These exclusions keep every entry reproducible, but leave out real
vulnerabilities. Studying how such vulnerabilities are exploited and
patched remains an important research direction.
For each advisory, \system\ includes at least one accepted
PoC, but does not enumerate every possible exploit format by 
combining vulnerable entry points, sinks, payloads, and paths.
Likewise, a patched version that blocks the PoC provides a negative case for that
particular attack, not proof that every variant has been eliminated.

\paragraph{Threats to validity.}
Each PoC demonstrates a single observed execution. Its source--sink trace
depends on instrumentation coverage, taint propagation, and the sink
catalog. Code outside the instrumentation's visibility or a missed flow
can therefore cause a valid exploit to be rejected.

LLM-generated PoCs may place code in the wrong lifecycle phase.
We check that the claimed effect is absent before \texttt{exploit},
rejecting PoCs whose \texttt{setup} has already produced it.
However, legitimate setup code placed in \texttt{exploit} may use
forbidden capabilities, causing an otherwise valid exploit to be rejected.
We deliberately accept this loss of coverage to reduce the risk of
including invalid exploits.

\paragraph{Actionable findings.}
Recent entries involve larger packages, longer setup phases, and greater
executed-code coverage. These PoCs are absent from earlier datasets such
as \vulcan~\cite{brito2023study} and \secbench~\cite{bhuiyan2023secbench}, suggesting that evaluations limited to those
datasets may underrepresent the challenges faced by current tools.
Evaluations should therefore include recent PoCs, report performance by
publication year, package, and vulnerability class, and use the supplied
traces to investigate failures. For repair tools, the patched version
provides a reference for assessing whether a proposed patch blocks the
demonstrated PoC.

%% file: Sections/Related_Work.tex
\section{Related Work}
\label{sec:relatedWork}

\runin{Vulnerability datasets}
Numerous vulnerability datasets offer different forms of ground truth.
Juliet~\cite{black2018juliet} supports controlled tool comparisons with small, synthetic C/C++ and Java programs containing known weaknesses.
Big-Vul~\cite{fan2020ac} and DiverseVul~\cite{chen2023diversevul} instead mine real C/C++ vulnerabilities and their patches to support learning-based detection.
PrimeVul~\cite{primevul} addresses label quality and duplication, showing how dataset construction can substantially affect conclusions about detection tools.
CVEfixes~\cite{bhandari2021cvefixes} extends automated collection across programming languages, linking CVE records to vulnerable code and fixing commits through a repeatable pipeline.
These resources document vulnerable code at scale, but not how to trigger it.

Executable benchmarks close this gap with reproducible tests.
For Java, Vul4J~\cite{bui2022vul4j} pairs vulnerabilities with patches and proof-of-vulnerability tests that distinguish the vulnerable and patched versions.
For Node.js, \vulcan~\cite{brito2023study} and \secbench~\cite{bhuiyan2023secbench} are the closest to our work, focusing on vulnerability locations and executable exploits, respectively.
We reuse some of their advisories and artifacts (\S\ref{subsec:collection}) and compare both datasets with \system in \S\ref{sec:benchmark}.
\system combines their executable exploits and source and sink locations, adding execution traces, patches checked against exploits, and infrastructure for benchmark growth and entry reproducibility.

\runin{Vulnerability analysis for Node.js}
Many tools detect or mitigate taint-style vulnerabilities in Node.js packages.
ODGen~\cite{li2022odgen} and Graph.js~\cite{ferreira2024graphjs} build and statically query package graphs for flows from exported functions to sensitive sinks; FAST~\cite{kang2023FAST} scales abstract interpretation to detect these flows and also produces exploits.
Dynamic analyses observe executions instead, as NodeMedic~\cite{cassel2023nodemedic} does by tracking provenance from package inputs to sinks.
Silent Spring~\cite{shcherbakov2023silent} and GHunter~\cite{cornelissen2024ghunter} target prototype pollution, searching for gadgets that turn it into code execution.
Synode~\cite{staicu2018synode} combines static analysis with runtime enforcement to prevent injection attacks.
These tools have been evaluated on known vulnerabilities, often from \vulcan and \secbench; \system offers a larger, more recent collection.

\runin{Automated exploit generation}
Generating an exploit often requires reconstructing how to reach a vulnerability through a package's interface.
Explode.js~\cite{marques2025automated} combines static analysis and symbolic execution to synthesize exploits, including those requiring multiple API interactions.
NodeMedic-FINE~\cite{cassel2025nodemedicfine} combines dynamic taint analysis, fuzzing, and exploit synthesis.
Both produce executable witnesses and benefit from benchmarks of known exploits for their evaluation.
PoCGen~\cite{simsek2026pocgen} generates working exploits from informal advisories using LLMs prompted with vulnerability reports, static taint paths, and codebase usage examples, refining candidates through runtime feedback.
A candidate is accepted if it passes a class-specific effect check, if the vulnerable function appears on the call stack when the effect occurs, and if an LLM judges that it triggers the reported vulnerability; for code injection, a static check also rejects exploits that call the sink themselves.
PoCGen produces working exploits for 71\% of \secbench vulnerabilities and 60\% of 126 recent advisories, supporting LLM-based generation for benchmark growth as vulnerabilities are reported.
Our pipeline uses the same generate--validate--refine pattern to produce benchmark entries rather than individual exploits, supporting repeated execution, analysis, and patch checking.
Lifecycle phases explicitly separate legitimate preparation from manufacturing the claimed effect.
Only program analyses validate exploits; acceptance never depends on an LLM's judgement.
The same analyses recover the source and sink locations and the trace stored with each entry.
PoCGen exploits could also enter \system after lifecycle translation and validation under the same rules.

\runin{AI-assisted benchmark construction}
LLMs increasingly support benchmark construction.
FormAI~\cite{tihanyi2023formai} generates C programs with an LLM and labels their vulnerabilities using bounded model checking.
\textsc{SeCodePLT}~\cite{nie2026secodeplt} expands validated seed examples through LLM-based mutations, producing security tasks in Python, C/C++, and Java with tests, vulnerable and patched code, and PoCs.
Both rely on automated checks to corroborate the generated data.
\system's vulnerable code comes from published npm packages; the LLM generates only exploits for the targeted vulnerabilities.

Other recent benchmarks evaluate LLM agents on real vulnerabilities rather than generated code: CyberGym~\cite{wang2026cybergym} asks agents to reproduce OSS-Fuzz vulnerabilities from descriptions; BountyBench~\cite{zhang2026bountybench} turns bug-bounty reports into detection, exploitation, and patching tasks in real systems.
SEC-bench~\cite{lee2025secbench} is our closest precedent for constructing benchmarks with LLM agents.
Its agents construct execution environments, reproduce real vulnerabilities, and produce patches, yielding a benchmark for PoC generation and vulnerability repair.
It targets C/C++ vulnerabilities checked through sanitizer reports.
Like SEC-bench, \system accepts an exploit only when a program analysis confirms it, never on an LLM's judgement. Unlike memory-safety bugs, however, taint-style vulnerabilities in Node.js lack a ready-made oracle such as a sanitizer.
\system therefore builds one from dynamic taint tracking and class-specific checks on the exploit's effect (\S\ref{subsec:validation}).

\runin{Live benchmarks}
Static benchmarks become outdated and may enter the training data of LLMs subsequently evaluated on them.
Live benchmarks address both problems by continuously adding new data.
LiveCodeBench~\cite{jain2025livecodebench} collects newly published programming problems from contest platforms, enabling evaluation on problems released after models' training cutoffs.
SWE-bench-Live~\cite{zhang2026swe} applies this approach to issue resolution, adding newly created GitHub issues through automated curation.
\system's pipeline applies this approach to vulnerabilities, turning newly published advisories into validated entries for evaluating tools on newly added vulnerabilities and exploits.

%% file: Sections/Conclusion.tex
\section{Conclusions and Future Work}
\label{sec:conclusionsAndFutureWork}

We presented \system, a live benchmark of Node.js taint-style vulnerabilities in four classes: code injection, OS command injection, prototype pollution, and path traversal. Its current release contains \nValidExploits validated exploits for \nValidPackages packages, each in a lifecycle format that separates the attack from its harness and, when one exists, paired with a patch checked against the exploit. Acceptance never depends on an LLM's judgement: program analyses check that an exploit carries attacker input from a source to a sink of the reported class and produces a confirmed effect, without a capability that produces that effect by itself. The same analyses record the source, sink, and execution trace of each validated exploit, and adaptive instrumentation lets them scale to large packages. A self-hosted mirror of every required package version keeps entries executable, and the pipeline turns new advisories into new entries.

Using \system, we found that these vulnerabilities have changed considerably since 2022, when the most widely used datasets end. Counting each package once, the median vulnerable package grew from about \medPkgSlocEarly lines of code in 2021--2022 to \medPkgSlocLate in 2026, and exploits need several times more setup code. Almost half of the 2026 advisories concern a vulnerability class already patched in the same package. Patch checking also revealed two widely used packages still exploitable in their latest releases, which their advisories list as patched (\S\ref{subsec:trend-zerodays}).

Several directions remain open. First, regular releases will let us track the trends of \S\ref{sec:metrics}, in particular the rise in recurrences and the possible role of AI-assisted discovery and development. Second, we plan to extend \system\ to further vulnerability classes, to advisory sources beyond the GitHub Advisory Database, and to vulnerabilities whose sink lies in a consuming application, currently excluded (\S\ref{subsec:exclusions}) because they require pairing a package with a client. Third, validation can be strengthened by reducing the manual inspection of exploits whose capabilities have ambiguous intent, by covering more entry points and payloads per vulnerability, and by testing every patched version against variants of its exploit to find incomplete patches. Finally, we intend to use \system\ to evaluate existing Node.js analyses on recent vulnerabilities in large packages.

%% file: main.bbl

\begin{thebibliography}{53}

%%% the Web address.

\ifx \showCODEN    \undefined \def \showCODEN     #1{\unskip}     \fi
\ifx \showDOI      \undefined \def \showDOI       #1{#1}\fi
\ifx \showISBNx    \undefined \def \showISBNx     #1{\unskip}     \fi
\ifx \showISBNxiii \undefined \def \showISBNxiii  #1{\unskip}     \fi
\ifx \showISSN     \undefined \def \showISSN      #1{\unskip}     \fi
\ifx \showLCCN     \undefined \def \showLCCN      #1{\unskip}     \fi
\ifx \shownote     \undefined \def \shownote      #1{#1}          \fi
\ifx \showarticletitle \undefined \def \showarticletitle #1{#1}   \fi
\ifx \showURL      \undefined \def \showURL       {\relax}        \fi
% invisible to TeX
\providecommand\bibfield[2]{#2}
\providecommand\bibinfo[2]{#2}
\providecommand\natexlab[1]{#1}
\providecommand\showeprint[2][]{arXiv:#2}

\bibitem[ver(2015)]%
        {verdaccio}
 \bibinfo{year}{2015}\natexlab{}.
\newblock \bibinfo{title}{Verdaccio}.
\newblock
\newblock
\urldef\tempurl%
\url{https://www.verdaccio.org/}
\showURL{%
\tempurl}
\newblock
\shownote{Accessed: 2026}.


\bibitem[{Anthropic}(2026)]%
        {anthropic2026firefox}
\bibfield{author}{\bibinfo{person}{{Anthropic}}.}
  \bibinfo{year}{2026}\natexlab{}.
\newblock \bibinfo{title}{Partnering with {Mozilla} to Improve {Firefox}'s
  Security}.
\newblock
\newblock
\urldef\tempurl%
\url{https://anthropic.com/news/mozilla-firefox-security}
\showURL{%
\tempurl}
\newblock
\shownote{Accessed: 2026-09-29}.


\bibitem[Baker et~al\mbox{.}(2025)]%
        {baker2025monitoring}
\bibfield{author}{\bibinfo{person}{Bowen Baker}, \bibinfo{person}{Joost
  Huizinga}, \bibinfo{person}{Leo Gao}, \bibinfo{person}{Zehao Dou},
  \bibinfo{person}{Melody~Y. Guan}, \bibinfo{person}{Aleksander Madry},
  \bibinfo{person}{Wojciech Zaremba}, \bibinfo{person}{Jakub Pachocki}, {and}
  \bibinfo{person}{David Farhi}.} \bibinfo{year}{2025}\natexlab{}.
\newblock \showarticletitle{Monitoring Reasoning Models for Misbehavior and the
  Risks of Promoting Obfuscation}.
\newblock \bibinfo{journal}{\emph{arXiv preprint arXiv:2503.11926v1}}
  (\bibinfo{year}{2025}).
\newblock


\bibitem[Bhandari et~al\mbox{.}(2021)]%
        {bhandari2021cvefixes}
\bibfield{author}{\bibinfo{person}{Guru Bhandari}, \bibinfo{person}{Amara
  Naseer}, {and} \bibinfo{person}{Leon Moonen}.}
  \bibinfo{year}{2021}\natexlab{}.
\newblock \showarticletitle{CVEfixes: automated collection of vulnerabilities
  and their fixes from open-source software}. In
  \bibinfo{booktitle}{\emph{Proceedings of the 17th international conference on
  predictive models and data analytics in software engineering}}.
  \bibinfo{pages}{30--39}.
\newblock


\bibitem[Bhuiyan et~al\mbox{.}(2023)]%
        {bhuiyan2023secbench}
\bibfield{author}{\bibinfo{person}{Masudul Hasan~Masud Bhuiyan},
  \bibinfo{person}{Adithya~Srinivas Parthasarathy}, \bibinfo{person}{Nikos
  Vasilakis}, \bibinfo{person}{Michael Pradel}, {and}
  \bibinfo{person}{Cristian-Alexandru Staicu}.}
  \bibinfo{year}{2023}\natexlab{}.
\newblock \showarticletitle{Secbench. js: An executable security benchmark
  suite for server-side javascript}. In \bibinfo{booktitle}{\emph{2023 IEEE/ACM
  45th International Conference on Software Engineering (ICSE)}}. IEEE,
  \bibinfo{pages}{1059--1070}.
\newblock


\bibitem[Black(2018)]%
        {black2018juliet}
\bibfield{author}{\bibinfo{person}{Paul~E Black}.}
  \bibinfo{year}{2018}\natexlab{}.
\newblock \bibinfo{booktitle}{\emph{Juliet 1.3 test suite: Changes from 1.2}}.
\newblock \bibinfo{publisher}{US Department of Commerce, National Institute of
  Standards and Technology~…}.
\newblock


\bibitem[Brito et~al\mbox{.}(2023)]%
        {brito2023study}
\bibfield{author}{\bibinfo{person}{Tiago Brito}, \bibinfo{person}{Mafalda
  Ferreira}, \bibinfo{person}{Miguel Monteiro}, \bibinfo{person}{Pedro Lopes},
  \bibinfo{person}{Miguel Barros}, \bibinfo{person}{Jos{\'e}~Fragoso Santos},
  {and} \bibinfo{person}{Nuno Santos}.} \bibinfo{year}{2023}\natexlab{}.
\newblock \showarticletitle{Study of javascript static analysis tools for
  vulnerability detection in node. js packages}.
\newblock \bibinfo{journal}{\emph{IEEE Transactions on Reliability}}
  \bibinfo{volume}{72}, \bibinfo{number}{4} (\bibinfo{year}{2023}),
  \bibinfo{pages}{1324--1339}.
\newblock


\bibitem[Bui et~al\mbox{.}(2022)]%
        {bui2022vul4j}
\bibfield{author}{\bibinfo{person}{Quang-Cuong Bui}, \bibinfo{person}{Riccardo
  Scandariato}, {and} \bibinfo{person}{Nicol{\'a}s E~D{\'\i}az Ferreyra}.}
  \bibinfo{year}{2022}\natexlab{}.
\newblock \showarticletitle{Vul4j: A dataset of reproducible java
  vulnerabilities geared towards the study of program repair techniques}. In
  \bibinfo{booktitle}{\emph{Proceedings of the 19th International Conference on
  Mining Software Repositories}}. \bibinfo{pages}{464--468}.
\newblock


\bibitem[Cassel et~al\mbox{.}(2025)]%
        {cassel2025nodemedicfine}
\bibfield{author}{\bibinfo{person}{Darion Cassel}, \bibinfo{person}{Nuno
  Sabino}, \bibinfo{person}{Min-Chien Hsu}, \bibinfo{person}{Ruben Martins},
  {and} \bibinfo{person}{Limin Jia}.} \bibinfo{year}{2025}\natexlab{}.
\newblock \showarticletitle{NodeMedic-FINE: Automatic Detection and Exploit
  Synthesis for Node. js Vulnerabilities}. In \bibinfo{booktitle}{\emph{NDSS}}.
\newblock


\bibitem[Cassel et~al\mbox{.}(2023)]%
        {cassel2023nodemedic}
\bibfield{author}{\bibinfo{person}{Darion Cassel}, \bibinfo{person}{Wai~Tuck
  Wong}, {and} \bibinfo{person}{Limin Jia}.} \bibinfo{year}{2023}\natexlab{}.
\newblock \showarticletitle{Nodemedic: End-to-end analysis of node. js
  vulnerabilities with provenance graphs}. In \bibinfo{booktitle}{\emph{2023
  IEEE 8th European Symposium on Security and Privacy (EuroS\&P)}}. IEEE,
  \bibinfo{pages}{1101--1127}.
\newblock


\bibitem[Chen et~al\mbox{.}(2026)]%
        {chen2026securevibebench}
\bibfield{author}{\bibinfo{person}{Junkai Chen}, \bibinfo{person}{Huihui
  Huang}, \bibinfo{person}{Yunbo Lyu}, \bibinfo{person}{Junwen An},
  \bibinfo{person}{Jieke Shi}, \bibinfo{person}{Chengran Yang},
  \bibinfo{person}{Ting Zhang}, \bibinfo{person}{Haoye Tian},
  \bibinfo{person}{Yikun Li}, \bibinfo{person}{Zhenhao Li},
  \bibinfo{person}{Xin Zhou}, \bibinfo{person}{Xing Hu}, {and}
  \bibinfo{person}{David Lo}.} \bibinfo{year}{2026}\natexlab{}.
\newblock \showarticletitle{{SecureVibeBench}: Benchmarking Secure Vibe Coding
  of {AI} Agents via Reconstructing Vulnerability-Introducing Scenarios}. In
  \bibinfo{booktitle}{\emph{Proceedings of the Annual Meeting of the
  Association for Computational Linguistics (ACL)}}.
\newblock


\bibitem[Chen et~al\mbox{.}(2023)]%
        {chen2023diversevul}
\bibfield{author}{\bibinfo{person}{Yizheng Chen}, \bibinfo{person}{Zhoujie
  Ding}, \bibinfo{person}{Lamya Alowain}, \bibinfo{person}{Xinyun Chen}, {and}
  \bibinfo{person}{David Wagner}.} \bibinfo{year}{2023}\natexlab{}.
\newblock \showarticletitle{Diversevul: A new vulnerable source code dataset
  for deep learning based vulnerability detection}. In
  \bibinfo{booktitle}{\emph{Proceedings of the 26th international symposium on
  research in attacks, intrusions and defenses}}. \bibinfo{pages}{654--668}.
\newblock


\bibitem[Cornelissen et~al\mbox{.}(2024)]%
        {cornelissen2024ghunter}
\bibfield{author}{\bibinfo{person}{Eric Cornelissen}, \bibinfo{person}{Mikhail
  Shcherbakov}, {and} \bibinfo{person}{Musard Balliu}.}
  \bibinfo{year}{2024}\natexlab{}.
\newblock \showarticletitle{GHunter: Universal Prototype Pollution Gadgets in
  JavaScript Runtimes}. In \bibinfo{booktitle}{\emph{33rd USENIX Security
  Symposium (USENIX Security 24)}}. \bibinfo{pages}{3693--3710}.
\newblock


\bibitem[Ding et~al\mbox{.}(2025)]%
        {primevul}
\bibfield{author}{\bibinfo{person}{Yangruibo Ding}, \bibinfo{person}{Yanjun
  Fu}, \bibinfo{person}{Omniyyah Ibrahim}, \bibinfo{person}{Chawin Sitawarin},
  \bibinfo{person}{Xinyun Chen}, \bibinfo{person}{Basel Alomair},
  \bibinfo{person}{David Wagner}, \bibinfo{person}{Baishakhi Ray}, {and}
  \bibinfo{person}{Yizheng Chen}.} \bibinfo{year}{2025}\natexlab{}.
\newblock \showarticletitle{Vulnerability Detection with Code Language Models:
  How Far are We?}. In \bibinfo{booktitle}{\emph{2025 IEEE/ACM 47th
  International Conference on Software Engineering (ICSE)}}.
  \bibinfo{pages}{1729--1741}.
\newblock
\urldef\tempurl%
\url{https://doi.org/10.1109/ICSE55347.2025.00038}
\showDOI{\tempurl}


\bibitem[{Discord Inc.}(2015)]%
        {discord}
\bibfield{author}{\bibinfo{person}{{Discord Inc.}}}
  \bibinfo{year}{2015}\natexlab{}.
\newblock \bibinfo{title}{Discord}.
\newblock
\newblock
\urldef\tempurl%
\url{https://discord.com}
\showURL{%
\tempurl}
\newblock
\shownote{Accessed: 2024-06-10}.


\bibitem[Du et~al\mbox{.}(2026)]%
        {vulrag}
\bibfield{author}{\bibinfo{person}{Xueying Du}, \bibinfo{person}{Geng Zheng},
  \bibinfo{person}{Kaixin Wang}, \bibinfo{person}{Yi Zou},
  \bibinfo{person}{Yujia Wang}, \bibinfo{person}{Wentai Deng},
  \bibinfo{person}{Jiayi Feng}, \bibinfo{person}{Mingwei Liu},
  \bibinfo{person}{Bihuan Chen}, \bibinfo{person}{Xin Peng},
  \bibinfo{person}{Tao Ma}, {and} \bibinfo{person}{Yiling Lou}.}
  \bibinfo{year}{2026}\natexlab{}.
\newblock \showarticletitle{Vul-rag: Enhancing llm-based vulnerability
  detection via knowledge-level rag}.
\newblock \bibinfo{journal}{\emph{ACM Transactions on Software Engineering and
  Methodology}} \bibinfo{volume}{35}, \bibinfo{number}{10}, Article
  \bibinfo{articleno}{312} (\bibinfo{year}{2026}),
  \bibinfo{numpages}{25}~pages.
\newblock
\urldef\tempurl%
\url{https://doi.org/10.1145/3797277}
\showDOI{\tempurl}


\bibitem[Fan et~al\mbox{.}(2020)]%
        {fan2020ac}
\bibfield{author}{\bibinfo{person}{Jiahao Fan}, \bibinfo{person}{Yi Li},
  \bibinfo{person}{Shaohua Wang}, {and} \bibinfo{person}{Tien~N Nguyen}.}
  \bibinfo{year}{2020}\natexlab{}.
\newblock \showarticletitle{A C/C++ code vulnerability dataset with code
  changes and CVE summaries}. In \bibinfo{booktitle}{\emph{Proceedings of the
  17th international conference on mining software repositories}}.
  \bibinfo{pages}{508--512}.
\newblock


\bibitem[Ferreira et~al\mbox{.}(2024)]%
        {ferreira2024graphjs}
\bibfield{author}{\bibinfo{person}{Mafalda Ferreira}, \bibinfo{person}{Miguel
  Monteiro}, \bibinfo{person}{Tiago Brito}, \bibinfo{person}{Miguel~E Coimbra},
  \bibinfo{person}{Nuno Santos}, \bibinfo{person}{Limin Jia}, {and}
  \bibinfo{person}{Jos{\'e}~Fragoso Santos}.} \bibinfo{year}{2024}\natexlab{}.
\newblock \showarticletitle{Efficient static vulnerability analysis for
  javascript with multiversion dependency graphs}.
\newblock \bibinfo{journal}{\emph{Proceedings of the ACM on Programming
  Languages}} \bibinfo{volume}{8}, \bibinfo{number}{PLDI}
  (\bibinfo{year}{2024}), \bibinfo{pages}{417--441}.
\newblock


\bibitem[{GitHub, Inc.}(2026a)]%
        {ghsa}
\bibfield{author}{\bibinfo{person}{{GitHub, Inc.}}}
  \bibinfo{year}{2026}\natexlab{a}.
\newblock \bibinfo{title}{GitHub Advisory Database}.
\newblock \bibinfo{howpublished}{\url{https://github.com/advisories}}.
\newblock
\newblock
\shownote{Accessed: 2024-06-10}.


\bibitem[{GitHub, Inc.}(2026b)]%
        {ghsa-docs}
\bibfield{author}{\bibinfo{person}{{GitHub, Inc.}}}
  \bibinfo{year}{2026}\natexlab{b}.
\newblock \bibinfo{title}{GitHub Advisory database - GitHub Docs}.
\newblock
\newblock
\urldef\tempurl%
\url{https://docs.github.com/en/code-security/concepts/vulnerability-reporting-and-management/github-advisory-database}
\showURL{%
\tempurl}
\newblock
\shownote{Accessed: 2026-09-1}.


\bibitem[{Google LLC}(2026)]%
        {osv}
\bibfield{author}{\bibinfo{person}{{Google LLC}}.}
  \bibinfo{year}{2026}\natexlab{}.
\newblock \bibinfo{title}{OSV: Open Source Vulnerabilities}.
\newblock \bibinfo{howpublished}{\url{https://osv.dev}}.
\newblock
\newblock
\shownote{Accessed: 2024-06-10}.


\bibitem[He et~al\mbox{.}(2026)]%
        {he2026speed}
\bibfield{author}{\bibinfo{person}{Hao He}, \bibinfo{person}{Courtney Miller},
  \bibinfo{person}{Shyam Agarwal}, \bibinfo{person}{Christian K{\"a}stner},
  {and} \bibinfo{person}{Bogdan Vasilescu}.} \bibinfo{year}{2026}\natexlab{}.
\newblock \showarticletitle{Speed at the cost of quality: How Cursor AI
  increases short-term velocity and long-term complexity in open-source
  projects}. In \bibinfo{booktitle}{\emph{Proceedings of the 23rd international
  conference on mining software repositories}}. \bibinfo{pages}{181--193}.
\newblock


\bibitem[Jain et~al\mbox{.}(2025)]%
        {jain2025livecodebench}
\bibfield{author}{\bibinfo{person}{Naman Jain}, \bibinfo{person}{Alex Gu},
  \bibinfo{person}{Wen-Ding Li}, \bibinfo{person}{Fanjia Yan},
  \bibinfo{person}{Tianjun Zhang}, \bibinfo{person}{Sida Wang},
  \bibinfo{person}{Armando Solar-Lezama}, \bibinfo{person}{Koushik Sen}, {and}
  \bibinfo{person}{Ion Stoica}.} \bibinfo{year}{2025}\natexlab{}.
\newblock \showarticletitle{Livecodebench: Holistic and contamination free
  evaluation of large language models for code}. In
  \bibinfo{booktitle}{\emph{International Conference on Learning
  Representations}}, Vol.~\bibinfo{volume}{2025}.
  \bibinfo{pages}{58791--58831}.
\newblock


\bibitem[Kang et~al\mbox{.}(2023)]%
        {kang2023FAST}
\bibfield{author}{\bibinfo{person}{Mingqing Kang}, \bibinfo{person}{Yichao Xu},
  \bibinfo{person}{Song Li}, \bibinfo{person}{Rigel Gjomemo},
  \bibinfo{person}{Jianwei Hou}, \bibinfo{person}{VN Venkatakrishnan}, {and}
  \bibinfo{person}{Yinzhi Cao}.} \bibinfo{year}{2023}\natexlab{}.
\newblock \showarticletitle{Scaling javascript abstract interpretation to
  detect and exploit node. js taint-style vulnerability}. In
  \bibinfo{booktitle}{\emph{2023 IEEE Symposium on Security and Privacy (SP)}}.
  IEEE, \bibinfo{pages}{1059--1076}.
\newblock


\bibitem[Lee et~al\mbox{.}(2025)]%
        {lee2025secbench}
\bibfield{author}{\bibinfo{person}{Hwiwon Lee}, \bibinfo{person}{Ziqi Zhang},
  \bibinfo{person}{Hanxiao Lu}, {and} \bibinfo{person}{Lingming Zhang}.}
  \bibinfo{year}{2025}\natexlab{}.
\newblock \showarticletitle{{SEC-bench}: Automated Benchmarking of {LLM} Agents
  on Real-World Software Security Tasks}. In \bibinfo{booktitle}{\emph{Advances
  in Neural Information Processing Systems}}, Vol.~\bibinfo{volume}{38}.
\newblock
\urldef\tempurl%
\url{https://doi.org/10.52202/085713-3878}
\showDOI{\tempurl}


\bibitem[Li et~al\mbox{.}(2022)]%
        {li2022odgen}
\bibfield{author}{\bibinfo{person}{Song Li}, \bibinfo{person}{Mingqing Kang},
  \bibinfo{person}{Jianwei Hou}, {and} \bibinfo{person}{Yinzhi Cao}.}
  \bibinfo{year}{2022}\natexlab{}.
\newblock \showarticletitle{Mining node. js vulnerabilities via object
  dependence graph and query}. In \bibinfo{booktitle}{\emph{31st USENIX
  Security Symposium (USENIX Security 22)}}. \bibinfo{pages}{143--160}.
\newblock


\bibitem[Li et~al\mbox{.}(2025)]%
        {li2025iris}
\bibfield{author}{\bibinfo{person}{Ziyang Li}, \bibinfo{person}{Saikat Dutta},
  {and} \bibinfo{person}{Mayur Naik}.} \bibinfo{year}{2025}\natexlab{}.
\newblock \showarticletitle{IRIS: LLM-assisted static analysis for detecting
  security vulnerabilities}. In \bibinfo{booktitle}{\emph{International
  conference on learning representations}}, Vol.~\bibinfo{volume}{2025}.
  \bibinfo{pages}{35735--35758}.
\newblock


\bibitem[Liu et~al\mbox{.}(2022)]%
        {liu2022demystifying}
\bibfield{author}{\bibinfo{person}{Chengwei Liu}, \bibinfo{person}{Sen Chen},
  \bibinfo{person}{Lingling Fan}, \bibinfo{person}{Bihuan Chen},
  \bibinfo{person}{Yang Liu}, {and} \bibinfo{person}{Xin Peng}.}
  \bibinfo{year}{2022}\natexlab{}.
\newblock \showarticletitle{Demystifying the vulnerability propagation and its
  evolution via dependency trees in the npm ecosystem}. In
  \bibinfo{booktitle}{\emph{Proceedings of the 44th international conference on
  software engineering}}. \bibinfo{pages}{672--684}.
\newblock


\bibitem[Marques et~al\mbox{.}(2025)]%
        {marques2025automated}
\bibfield{author}{\bibinfo{person}{Filipe Marques}, \bibinfo{person}{Mafalda
  Ferreira}, \bibinfo{person}{Andr{\'e} Nascimento}, \bibinfo{person}{Miguel~E
  Coimbra}, \bibinfo{person}{Nuno Santos}, \bibinfo{person}{Limin Jia}, {and}
  \bibinfo{person}{Jos{\'e} Fragoso~Santos}.} \bibinfo{year}{2025}\natexlab{}.
\newblock \showarticletitle{Automated exploit generation for node. js
  packages}.
\newblock \bibinfo{journal}{\emph{Proceedings of the ACM on Programming
  Languages}} \bibinfo{volume}{9}, \bibinfo{number}{PLDI}
  (\bibinfo{year}{2025}), \bibinfo{pages}{1341--1366}.
\newblock


\bibitem[Microsoft(2015)]%
        {vscode}
\bibfield{author}{\bibinfo{person}{Microsoft}.}
  \bibinfo{year}{2015}\natexlab{}.
\newblock \bibinfo{title}{Visual Studio Code}.
\newblock
\newblock
\urldef\tempurl%
\url{https://github.com/microsoft/vscode}
\showURL{%
\tempurl}
\newblock
\shownote{Accessed: 2025}.


\bibitem[Nguyen et~al\mbox{.}(2025)]%
        {nguyen2025safe}
\bibfield{author}{\bibinfo{person}{Van Nguyen}, \bibinfo{person}{Surya Nepal},
  \bibinfo{person}{Xingliang Yuan}, \bibinfo{person}{Tingmin Wu}, {and}
  \bibinfo{person}{Carsten Rudolph}.} \bibinfo{year}{2025}\natexlab{}.
\newblock \showarticletitle{Safe: A novel approach for software vulnerability
  detection from enhancing the capability of large language models}. In
  \bibinfo{booktitle}{\emph{Proceedings of the 20th ACM Asia Conference on
  Computer and Communications Security}}. \bibinfo{pages}{392--406}.
\newblock
\urldef\tempurl%
\url{https://doi.org/10.1145/3708821.3736208}
\showDOI{\tempurl}


\bibitem[Ni et~al\mbox{.}(2026)]%
        {llmvdjs}
\bibfield{author}{\bibinfo{person}{Ronghao Ni}, \bibinfo{person}{Mihai
  Christodorescu}, {and} \bibinfo{person}{Limin Jia}.}
  \bibinfo{year}{2026}\natexlab{}.
\newblock \bibinfo{title}{Taint-Style Vulnerability Detection and Confirmation
  for Node.js Packages Using LLM Agent Reasoning}.
\newblock
\newblock
\urldef\tempurl%
\url{https://doi.org/10.48550/arXiv.2604.20179}
\showDOI{\tempurl}
\showeprint[arxiv]{2604.20179}~[cs.CR]


\bibitem[Nie et~al\mbox{.}(2026)]%
        {nie2026secodeplt}
\bibfield{author}{\bibinfo{person}{Yuzhou Nie}, \bibinfo{person}{Zhun Wang},
  \bibinfo{person}{Yu Yang}, \bibinfo{person}{Ruizhe Jiang},
  \bibinfo{person}{Yuheng Tang}, \bibinfo{person}{Xander Davies},
  \bibinfo{person}{Yarin Gal}, \bibinfo{person}{Bo Li}, \bibinfo{person}{Wenbo
  Guo}, {and} \bibinfo{person}{Dawn Song}.} \bibinfo{year}{2026}\natexlab{}.
\newblock \showarticletitle{SECODEPLT: A unified benchmark for evaluating the
  security risks and capabilities of code genAI}.
\newblock \bibinfo{journal}{\emph{Advances in Neural Information Processing
  Systems}}  \bibinfo{volume}{38} (\bibinfo{year}{2026}).
\newblock


\bibitem[{OpenClaw Foundation}(2026)]%
        {openclaw2026}
\bibfield{author}{\bibinfo{person}{{OpenClaw Foundation}}.}
  \bibinfo{year}{2026}\natexlab{}.
\newblock \bibinfo{booktitle}{\emph{OpenClaw}}.
\newblock
\urldef\tempurl%
\url{https://openclaw.ai}
\showURL{%
\tempurl}
\newblock
\shownote{Open-source AI assistant software}.


\bibitem[{OpenJS Foundation}(2025)]%
        {nodejs}
\bibfield{author}{\bibinfo{person}{{OpenJS Foundation}}.}
  \bibinfo{year}{2025}\natexlab{}.
\newblock \bibinfo{title}{Node.js}.
\newblock \bibinfo{howpublished}{\url{https://nodejs.org/}}.
\newblock
\newblock
\shownote{Version 20.17.0, accessed September 30, 2025}.


\bibitem[Pearce et~al\mbox{.}(2025)]%
        {pearce2025asleep}
\bibfield{author}{\bibinfo{person}{Hammond Pearce}, \bibinfo{person}{Baleegh
  Ahmad}, \bibinfo{person}{Benjamin Tan}, \bibinfo{person}{Brendan
  Dolan-Gavitt}, {and} \bibinfo{person}{Ramesh Karri}.}
  \bibinfo{year}{2025}\natexlab{}.
\newblock \showarticletitle{Asleep at the keyboard? assessing the security of
  github copilot’s code contributions}.
\newblock \bibinfo{journal}{\emph{Commun. ACM}} \bibinfo{volume}{68},
  \bibinfo{number}{2} (\bibinfo{year}{2025}), \bibinfo{pages}{96--105}.
\newblock


\bibitem[Perry et~al\mbox{.}(2023)]%
        {perry2023users}
\bibfield{author}{\bibinfo{person}{Neil Perry}, \bibinfo{person}{Megha
  Srivastava}, \bibinfo{person}{Deepak Kumar}, {and} \bibinfo{person}{Dan
  Boneh}.} \bibinfo{year}{2023}\natexlab{}.
\newblock \showarticletitle{Do users write more insecure code with ai
  assistants?}. In \bibinfo{booktitle}{\emph{Proceedings of the 2023 ACM SIGSAC
  conference on computer and communications security}}.
  \bibinfo{pages}{2785--2799}.
\newblock


\bibitem[Sen et~al\mbox{.}(2013)]%
        {jalangi-fse13}
\bibfield{author}{\bibinfo{person}{Koushik Sen}, \bibinfo{person}{Swaroop
  Kalasapur}, \bibinfo{person}{Tasneem Brutch}, {and} \bibinfo{person}{Simon
  Gibbs}.} \bibinfo{year}{2013}\natexlab{}.
\newblock \showarticletitle{Jalangi: a selective record-replay and dynamic
  analysis framework for {JavaScript}}. In
  \bibinfo{booktitle}{\emph{Proceedings of the 2013 9th Joint Meeting on
  Foundations of Software Engineering (ESEC/FSE)}}. \bibinfo{pages}{488–498}.
\newblock
\urldef\tempurl%
\url{https://doi.org/10.1145/2491411.2491447}
\showDOI{\tempurl}


\bibitem[Shcherbakov et~al\mbox{.}(2023)]%
        {shcherbakov2023silent}
\bibfield{author}{\bibinfo{person}{Mikhail Shcherbakov},
  \bibinfo{person}{Musard Balliu}, {and} \bibinfo{person}{Cristian-Alexandru
  Staicu}.} \bibinfo{year}{2023}\natexlab{}.
\newblock \showarticletitle{Silent spring: Prototype pollution leads to remote
  code execution in node. js}. In \bibinfo{booktitle}{\emph{32nd USENIX
  Security Symposium (USENIX Security 23)}}. \bibinfo{pages}{5521--5538}.
\newblock


\bibitem[Simsek et~al\mbox{.}(2026)]%
        {simsek2026pocgen}
\bibfield{author}{\bibinfo{person}{Deniz Simsek}, \bibinfo{person}{Aryaz
  Eghbali}, {and} \bibinfo{person}{Michael Pradel}.}
  \bibinfo{year}{2026}\natexlab{}.
\newblock \showarticletitle{Pocgen: Generating proof-of-concept exploits for
  vulnerabilities in npm packages}.
\newblock \bibinfo{journal}{\emph{Proceedings of the ACM on Software
  Engineering}} \bibinfo{volume}{3}, \bibinfo{number}{FSE}
  (\bibinfo{year}{2026}), \bibinfo{pages}{3887--3908}.
\newblock


\bibitem[{Slack Technologies}(2013)]%
        {slack}
\bibfield{author}{\bibinfo{person}{{Slack Technologies}}.}
  \bibinfo{year}{2013}\natexlab{}.
\newblock \bibinfo{title}{Slack}.
\newblock
\newblock
\urldef\tempurl%
\url{https://slack.com}
\showURL{%
\tempurl}
\newblock
\shownote{Accessed: 2024-06-10}.


\bibitem[{Snyk Ltd.}(2026)]%
        {snyk}
\bibfield{author}{\bibinfo{person}{{Snyk Ltd.}}}
  \bibinfo{year}{2026}\natexlab{}.
\newblock \bibinfo{title}{Snyk Vulnerability Database}.
\newblock \bibinfo{howpublished}{\url{https://security.snyk.io}}.
\newblock
\newblock
\shownote{Accessed: 2024-06-10}.


\bibitem[Staicu et~al\mbox{.}(2018)]%
        {staicu2018synode}
\bibfield{author}{\bibinfo{person}{Cristian-Alexandru Staicu},
  \bibinfo{person}{Michael Pradel}, {and} \bibinfo{person}{Ben Livshits}.}
  \bibinfo{year}{2018}\natexlab{}.
\newblock \showarticletitle{Understanding and automatically preventing
  injection attacks on Node. js}. In \bibinfo{booktitle}{\emph{Network and
  Distributed System Security Symposium (NDSS)}}.
\newblock


\bibitem[Tihanyi et~al\mbox{.}(2023)]%
        {tihanyi2023formai}
\bibfield{author}{\bibinfo{person}{Norbert Tihanyi}, \bibinfo{person}{Tamas
  Bisztray}, \bibinfo{person}{Ridhi Jain}, \bibinfo{person}{Mohamed~Amine
  Ferrag}, \bibinfo{person}{Lucas~C Cordeiro}, {and} \bibinfo{person}{Vasileios
  Mavroeidis}.} \bibinfo{year}{2023}\natexlab{}.
\newblock \showarticletitle{The formai dataset: Generative ai in software
  security through the lens of formal verification}. In
  \bibinfo{booktitle}{\emph{Proceedings of the 19th international conference on
  predictive models and data analytics in software engineering}}.
  \bibinfo{pages}{33--43}.
\newblock


\bibitem[Ullah et~al\mbox{.}(2024)]%
        {ullah2024llms}
\bibfield{author}{\bibinfo{person}{Saad Ullah}, \bibinfo{person}{Mingji Han},
  \bibinfo{person}{Saurabh Pujar}, \bibinfo{person}{Hammond Pearce},
  \bibinfo{person}{Ayse Coskun}, {and} \bibinfo{person}{Gianluca Stringhini}.}
  \bibinfo{year}{2024}\natexlab{}.
\newblock \showarticletitle{Llms cannot reliably identify and reason about
  security vulnerabilities (yet?): A comprehensive evaluation, framework, and
  benchmarks}. In \bibinfo{booktitle}{\emph{2024 IEEE symposium on security and
  privacy (SP)}}. IEEE, \bibinfo{pages}{862--880}.
\newblock
\urldef\tempurl%
\url{https://doi.org/10.1109/SP54263.2024.00210}
\showDOI{\tempurl}


\bibitem[University(2026)]%
        {dynajs}
\bibfield{author}{\bibinfo{person}{Programming Language Research Group @~Korea
  University}.} \bibinfo{year}{2026}\natexlab{}.
\newblock \bibinfo{title}{DynaJS}.
\newblock \bibinfo{howpublished}{\url{https://github.com/ku-plrg/dynajs}}.
\newblock


\bibitem[Wang et~al\mbox{.}(2026)]%
        {wang2026cybergym}
\bibfield{author}{\bibinfo{person}{Zhun Wang}, \bibinfo{person}{Tianneng Shi},
  \bibinfo{person}{Jingxuan He}, \bibinfo{person}{Matthew Cai},
  \bibinfo{person}{Jialin Zhang}, {and} \bibinfo{person}{Dawn Song}.}
  \bibinfo{year}{2026}\natexlab{}.
\newblock \showarticletitle{CyberGym: Evaluating AI Agents' Real-World
  Cybersecurity Capabilities at Scale}. In
  \bibinfo{booktitle}{\emph{International Conference on Learning
  Representations}}, Vol.~\bibinfo{volume}{2026}.
\newblock


\bibitem[Zhang et~al\mbox{.}(2025)]%
        {zhang2026bountybench}
\bibfield{author}{\bibinfo{person}{Andy~K. Zhang}, \bibinfo{person}{Joey Ji},
  \bibinfo{person}{Celeste Menders}, \bibinfo{person}{Riya Dulepet},
  \bibinfo{person}{Thomas Qin}, \bibinfo{person}{Ron~Y. Wang},
  \bibinfo{person}{Junrong Wu}, \bibinfo{person}{Kyleen Liao},
  \bibinfo{person}{Jiliang Li}, \bibinfo{person}{Jinghan Hu}, {et~al\mbox{.}}}
  \bibinfo{year}{2025}\natexlab{}.
\newblock \showarticletitle{BountyBench: Dollar Impact of AI Agent Attackers
  and Defenders on Real-World Cybersecurity Systems}. In
  \bibinfo{booktitle}{\emph{Advances in Neural Information Processing
  Systems}}, Vol.~\bibinfo{volume}{38}. \bibinfo{pages}{190575--190694}.
\newblock
\urldef\tempurl%
\url{https://doi.org/10.52202/085713-5725}
\showDOI{\tempurl}


\bibitem[Zhang et~al\mbox{.}(2026)]%
        {zhang2026swe}
\bibfield{author}{\bibinfo{person}{Linghao Zhang}, \bibinfo{person}{Shilin He},
  \bibinfo{person}{Chaoyun Zhang}, \bibinfo{person}{Yu Kang},
  \bibinfo{person}{Bowen Li}, \bibinfo{person}{Chengxing Xie},
  \bibinfo{person}{Junhao Wang}, \bibinfo{person}{Maoquan Wang},
  \bibinfo{person}{Yufan Huang}, \bibinfo{person}{Shengyu Fu}, {et~al\mbox{.}}}
  \bibinfo{year}{2026}\natexlab{}.
\newblock \showarticletitle{Swe-bench goes live!}
\newblock \bibinfo{journal}{\emph{Advances in Neural Information Processing
  Systems}}  \bibinfo{volume}{38} (\bibinfo{year}{2026}).
\newblock


\bibitem[Zhao et~al\mbox{.}(2025)]%
        {zhao2025vibe}
\bibfield{author}{\bibinfo{person}{Songwen Zhao}, \bibinfo{person}{Danqing
  Wang}, \bibinfo{person}{Kexun Zhang}, \bibinfo{person}{Jiaxuan Luo},
  \bibinfo{person}{Zhuo Li}, {and} \bibinfo{person}{Lei Li}.}
  \bibinfo{year}{2025}\natexlab{}.
\newblock \showarticletitle{Is vibe coding safe? Benchmarking vulnerability of
  agent-generated code in real-world tasks}.
\newblock \bibinfo{journal}{\emph{arXiv preprint arXiv:2512.03262}}
  (\bibinfo{year}{2025}).
\newblock


\bibitem[Zhong et~al\mbox{.}(2026)]%
        {zhong2026impossiblebench}
\bibfield{author}{\bibinfo{person}{Ziqian Zhong}, \bibinfo{person}{Aditi
  Raghunathan}, {and} \bibinfo{person}{Nicholas Carlini}.}
  \bibinfo{year}{2026}\natexlab{}.
\newblock \showarticletitle{ImpossibleBench: Measuring LLMs' Propensity of
  Exploiting Test Cases}. In \bibinfo{booktitle}{\emph{International Conference
  on Learning Representations}}, Vol.~\bibinfo{volume}{2026}.
  \bibinfo{pages}{124959--124986}.
\newblock


\bibitem[Zhu et~al\mbox{.}(2026)]%
        {10.1145/3808132}
\bibfield{author}{\bibinfo{person}{Hao Zhu}, \bibinfo{person}{Jia Li},
  \bibinfo{person}{Cuiyun Gao}, \bibinfo{person}{Jiaru Qian},
  \bibinfo{person}{Yihong Dong}, \bibinfo{person}{Huanyu Liu},
  \bibinfo{person}{Lecheng Wang}, \bibinfo{person}{Ziliang Wang},
  \bibinfo{person}{Xiaolong Hu}, {and} \bibinfo{person}{Ge Li}.}
  \bibinfo{year}{2026}\natexlab{}.
\newblock \showarticletitle{VulInstruct: Teaching LLMs Root-Cause Reasoning for
  Vulnerability Detection via Security Specifications}.
\newblock \bibinfo{journal}{\emph{Proc. ACM Softw. Eng.}} \bibinfo{volume}{3},
  \bibinfo{number}{FSE}, Article \bibinfo{articleno}{FSE125}
  (\bibinfo{date}{June} \bibinfo{year}{2026}), \bibinfo{numpages}{22}~pages.
\newblock
\urldef\tempurl%
\url{https://doi.org/10.1145/3808132}
\showDOI{\tempurl}


\bibitem[Zou(2026)]%
        {unit42_2026burst}
\bibfield{author}{\bibinfo{person}{Xu Zou}.} \bibinfo{year}{2026}\natexlab{}.
\newblock \bibinfo{title}{The Frontier {AI} Vulnerability Burst:
  Industrializing Autonomous Zero-Day Discovery in Open-Source Software}.
\newblock \bibinfo{howpublished}{Palo Alto Networks Unit 42}.
\newblock
\urldef\tempurl%
\url{https://unit42.paloaltonetworks.com/frontier-ai-vulnerability-burst/}
\showURL{%
\tempurl}
\newblock
\shownote{Accessed: 2026-09-29}.


\end{thebibliography}
